\documentclass[12pt]{article}
\usepackage{amsmath}
\usepackage{amssymb}
\usepackage{graphics}

\makeatletter\@addtoreset{equation}{section}
\makeatother
\usepackage[dvips]{color}
\allowdisplaybreaks[1]
\begin{document}
\begin{titlepage}
\begin{flushright}
TIT/HEP-615\\
KIAS-P11060\\
November 2011
\end{flushright}
\vspace{0.5cm}
\begin{center}
{\Large \bf
$\mathcal{N}=4$ Instanton Calculus in
$\Omega$ and R-R Backgrounds
}
\lineskip .75em
\vskip0.5cm
{\large Katsushi Ito${}^{1}$, Hiroaki Nakajima${}^{2,3}$, Takuya Saka${}^{1}$ and Shin
Sasaki${}^{4}$ }
\vskip 2.5em
${}^{1}$ {\normalsize\it Department of Physics,\\
Tokyo Institute of Technology\\
Tokyo, 152-8551, Japan} \vskip 1.0em
${}^{2}$ {\normalsize\it School of Physics,\\
Korea Institute for Advanced Study \\
Seoul, 130-722, Korea} \vskip 1.0em
${}^{3}$ {\normalsize\it Department of Physics,\\
Kyungpook National University \\
Taegu, 702-701, Korea} \vskip 1.0em
${}^{4}$ {\normalsize\it Department of Physics,\\
Kitasato University \\
Sagamihara, 228-8555, Japan}
\vskip 3.0em
\end{center}
\begin{abstract}
We study the instanton calculus for ${\cal N}=4$ super Yang-Mills 
theory in ten-dimensional $\Omega$-background with the R-symmetry 
Wilson line gauge field. 
{}From the ADHM construction of instantons in the background, 
we obtain the deformed instanton effective action. 
For a certain case 
we get the effective action of 
${\cal N}=2^*$ theory in the $\Omega$-background. 
We also study the low-energy effective D$(-1)$-brane 
action for the D3/D$(-1)$-brane system 
in the R-R 3-form field strength backgrounds 
and find that the action agrees with the instanton effective action 
in the $\Omega$-background. 
\end{abstract}
\end{titlepage}

\baselineskip=0.7cm
\tableofcontents
\section{Introduction}
The 
$\Omega$-background deformation 
\cite{Moore:1997dj} 
is used to perform the integrals over 
the moduli spaces in various
supersymmetric gauge theories via the localization technique. 
In particular, Nekrasov  computed the instanton partition 
function for ${\cal N}=2$ supersymmetric Yang-Mills theory 
from the $\Omega$-deformation \cite{Nekrasov:2002qd}, 
which is defined by the dimensional
reduction from the six-dimensional $\Omega$-background 
with the R-symmetry Wilson line gauge field~\cite{LoMaNe,Nekrasov:2003rj}.

It has been known that the
$\Omega$-background deformation can be 
interpreted 
as a 
certain ${\cal N}=2$ supergravity  background.
For the (anti-)self-dual $\Omega$-background,
the instanton partition function corresponds to the 
partition function of topological string theory
extracted from the scattering amplitudes including the self-dual 
graviphoton vertex operators in type II superstring 
theory~\cite{AnGaNaTa, BeCeOoVa}. 
The topological partition functions has been 
studied in the non-(anti-)self-dual case 
\cite{Awata:2005fa,Iqbal:2007ii,Huang:2010kf}.
Recently, it has been pointed out 
in \cite{Antoniadis:2010iq,Nakayama:2011be} that
the partition functions for the 
general (non-(anti-)self-dual) $\Omega$-background correspond to 
the scattering amplitudes 
including the anti-self-dual graviphoton 
and the self-dual gauge field 
associated with the matter vector multiplets. 

Such closed string  backgrounds also change the 
microscopic description of instantons realized by D-branes. 
It has been shown in \cite{Billo:2006jm} that the low-energy effective 
action of the D$(-1)$-branes 
for the D3/D$(-1)$-brane system 
at the fixed point of $\mathbb{C} \times {\mathbb
 C}^{2}/{\mathbb Z}_2$ in the self-dual R-R 3-form field strength 
 background 
coincides with the ${\cal N}=2$ 
instanton effective action in the 
self-dual $\Omega$-background.
In \cite{Ito:2010vx}, we have 
extended this
result to the general $\Omega$-background, 
where the corresponding closed string 
backgrounds are the (S,A)- and the (A,S)-types of 
the R-R 3-form field strengths.
The (S,A)-type R-R 3-form field strengths have the same tensor 
structures as the $\Omega$-backgrounds, which give the mass for the 
bosonic and the fermionic instanton moduli. 
On the other hand, the (A,S)-type field 
strengths correspond to the Wilson line gauge fields, which give the 
masses for the fermionic moduli. This correspondence is similar to 
the relation between the system in the uniform magnetic fields and 
the one in the rotating frame. 
%However precise meaning of this 
%correspondence is not clear at this moment.

%In order to confirm this relation in more general setup, we investigate 
%It would be interesting to explore 
In this work we will investigate
further generalization of the $\Omega$-background and its 
string theory description in order to understand this correspondence in
more general setup.
This is also useful for studying
%in order to study 
non-perturbative and stringy aspects of 
the supersymmetric gauge theories constructed by various D-branes. 
The $\Omega$-background deformation of D-brane systems in various dimensions 
has been studied in \cite{Jafferis:2007sg,Awata:2009dd,Ne-M,
Billo:2009di,Fucito:2009rs,Billo:2010mg}. 
In particular, we have proposed the ten-dimensional $\Omega$-background 
for the self-dual case and studied the deformed ${\cal N}=4$ super Yang-Mills 
theories \cite{Ito:2009ac}. 
We have studied the ADHM construction of instantons in the 
background and showed that the deformed instanton effective action 
agrees with the D$(-1)$-brane effective action 
for the D3/D$(-1)$ system in the 
self-dual 
R-R 3-form background of the (S,A)-type. 
We also found that the (A,S)-type R-R 3-form background gives the 
holomorphic mass deformation to the fermionic moduli in 
the instanton effective action.

In this paper we study the ${\cal N}=4$ action in the general
non-(anti-)self-dual $\Omega$-background with 
the R-symmetry Wilson line gauge field and 
the deformed instanton effective action. 
We also investigate the correspondence between the 
ten-dimensional $\Omega$-background and the R-R 3-form backgrounds. 
We will study the ADHM construction of instantons in the $\Omega$-deformed 
${\cal N}=4$ super Yang-Mills theory.
We find that this $\Omega$-deformation includes the mass 
deformation (${\cal N}=2^*$
deformation) as an example, which is useful
for the calculation of the ${\cal N}=4$
instanton partition function \cite{Hollowood:2002zv,Bruzzo:2002xf}.
We will calculate the D$(-1)$-brane effective action for 
the D3/D$(-1)$ system 
in the R-R 3-form backgrounds and find that the deformed action agrees with 
the instanton effective action in the 
ten-dimensional $\Omega$-background.

This paper is organized as follows:
in Section 2, we introduce the ten-dimensional 
$\Omega$-background and
the R-symmetry 
Wilson line 
gauge field.
We will discuss the ADHM construction of instantons in
the deformed ${\cal N}=4$ theory and calculate the instanton effective
action.
In Section 3 we will study the D$(-1)$-brane effective action 
for the D3/D$(-1)$ system in the 
R-R 3-form backgrounds.
Section 4 devotes the conclusions and discussion.
In Appendix A, we summarize our notations of the sigma matrices in four
and six dimensions.
In Appendix B, we describe 
detailed calculations of the disk amplitudes 
in the R-R 3-form backgrounds.

\section{$\mathcal{N}=4$ instanton effective action in $\Omega$-background}

\subsection{$\mathcal{N}=4$ super Yang-Mills theory in $\Omega$-background}
The $\mathcal{N}=4$ super Yang-Mills theory with the gauge group $U(N)$ 
contains a gauge field $A_m$ ($m=1,2,3,4$), 
Weyl fermions $\Lambda_{\alpha}^A, \bar{\Lambda}_{\dot{\alpha} A}$ 
($\alpha,\dot{\alpha}=1,2$, $A=1,2,3,4$) 
and six real scalar fields $\varphi_a$ ($a=1,\cdots,6$). 
The fields belong to the adjoint representation of the gauge group. 
Since we are interested in the instanton calculus, 
we study the theory defined in the Euclidean spacetime.
The left and the right spinor indices of the Lorentz 
group $SO(4) \cong SU(2)_L \times SU(2)_R$ are denoted by 
$\alpha$ and $\dot{\alpha}$ respectively.
They are raised and lowered 
by the anti-symmetric $\epsilon$-symbols 
$\epsilon_{\alpha\beta}$ and $\epsilon_{\dot{\alpha}\dot{\beta}}$
with $\epsilon^{12} = - \epsilon_{12} = 1$.
The index $A$ labels the (anti-)fundamental representation of the
R-symmetry group $SU(4)_I$.
The Lagrangian 
is 
\begin{align}
 \mathcal{L}_0 = \frac{1}{\kappa} \mathrm{Tr} \Big[ 
 &\, \frac{1}{4} F^{mn} F_{mn} + \frac{i \theta g^2}{32 \pi^2} F^{mn} \widetilde{F}_{mn} 
   + \Lambda^A \sigma^m D_m \bar{\Lambda}_A 
   + \frac{1}{2} \big( D_m \varphi_a \big)^2 \notag \\
 &\, - \frac{g}{2} ( \Sigma_a )^{AB} \bar{\Lambda}_A \big[ \varphi_a ,\bar{\Lambda}_B \big] 
   - \frac{g}{2} (\bar{\Sigma}_a)_{AB} \Lambda^A \big[ \varphi_a , \Lambda^B \big] 
   - \frac{g^2}{4} \big[ \varphi_a , \varphi_b \big]^2 \Big] ,
 \label{eq:Lag.undef}
\end{align}
where $g$ is the gauge coupling constant and $\theta$ is the theta angle
parameter.
$D_m * =  \partial_m * + ig \big[ A_m , * \big]$ is the gauge covariant
derivative,  $F_{mn} = \partial_m A_n - \partial_n A_m + ig \big[ A_m ,
A_n \big]$ is the gauge field strength and $\widetilde{F}_{mn} = \frac{1}{2}
\epsilon_{mnpq} F^{pq}$ is its dual.
We normalize the generators $T^u$ $(u=1,\cdots,N^2)$ of the gauge group $U(N)$ as $\mathrm{Tr} \big( T^u T^v \big) = \kappa \delta^{uv}$.
$\sigma^m$ and $\bar{\sigma}^m$ are the sigma matrices in four dimensions, 
 while $(\Sigma_a )^{AB}$ and $(\bar{\Sigma}_a)_{AB}$ are the sigma
 matrices in six dimensions.
We summarize their properties in Appendix~\ref{sec:sigma}.

The four-dimensional $\mathcal{N}=4$ super Yang-Mills theory is obtained 
by the dimensional reduction of the $\mathcal{N}=1$ super Yang-Mills 
theory in ten dimensions \cite{Brink:1976bc}. 
In this work, we consider the $\mathcal{N}=1$ super Yang-Mills theory
in the ten-dimensional $\Omega$-background and its reduction. 
We introduce the %{\bf curved [remove this word?]} 
ten-dimensional coordinates $x^{\mathcal{M}} = (x^m,x^{a+4})$ 
($\mathcal{M}=1,\cdots,10$) 
%[We should add this. However a=1,...,6. $x^a\rightarrow x^{a+4}$? 
%{\color{blue}How about $x^{a}\to y^{a}$?}]} 
and the metric $g_{\mathcal{M}\mathcal{N}}$ defined by 
\begin{align}
\mathrm{d}s^2=
g_{\mathcal{M}\mathcal{N}}\mathrm{d}x^{\mathcal{M}}\mathrm{d}x^{\mathcal{N}}=
\bigl(\mathrm{d}x^{a+4}\bigr)^2+
\bigl(\mathrm{d}x^{m}+\Omega^m_{a}\mathrm{d}x^{a+4}\bigr)^2 ,
\end{align}
with $\Omega^m_{a}\equiv \Omega^{mn}_{a}x_{n}$ 
where $\Omega_{mna}$ are six constant anti-symmetric matrices 
$\Omega^{mn}_a = - \Omega^{nm}_a$. 
The Lagrangian of the $\mathcal{N}=1$ super Yang-Mills theory 
in this background is 
\begin{align}
 \mathcal{L}_{10\text{D}} = \frac{1}{\kappa} \sqrt{- \det g_{\mathcal{M}\mathcal{N}}} \mathrm{Tr} \Big[ 
   - \frac{1}{4} g^{\mathcal{M}\mathcal{P}} g^{\mathcal{N}\mathcal{Q}} F_{\mathcal{M}\mathcal{N}} F_{\mathcal{P}\mathcal{Q}} 
   - \frac{i}{2} \bar{\Psi} e^{\mathcal{M}}{}_M \Gamma^M
 \mathcal{D}_{\mathcal{M}} \Psi \Big],
 \label{eq:10d.Lag}
\end{align}
where 
$F_{\mathcal{M}\mathcal{N}}  = \partial_{\mathcal{M}} A_{\mathcal{N}} 
 - \partial_{\mathcal{N}} A_{\mathcal{M}} + ig \big[ A_{\mathcal{M}} , A_{\mathcal{N}} \big] $ 
 is the field strength of the gauge field $A_{\mathcal{M}} = ( A_m , \varphi_a )$ 
 and 
 $\Psi$ is the ten-dimensional Majorana-Weyl spinor.
$\Gamma^M$ is the
 gamma matrices in ten dimensions and 
 $e^{\mathcal{M}}{}_M$ is the vielbein.
The vector indices in the local Lorentz frame is denoted by $M, N, \cdots$.
The covariant derivative in the curved spacetime for spinors is given by
\begin{align}
 \mathcal{D}_{\mathcal{M}} = D_{\mathcal{M}} - \frac{1}{2} \omega_{\mathcal{M}, MN} \Gamma^{MN} , 
 \label{eq:cov.deriv.sipnor}
\end{align}
where $D_{\mathcal{M}} * = \partial_{\mathcal{M}} * 
 + ig \big[ A_{\mathcal{M}} , * \big] $ is the gauge covariant derivative,
 $\Gamma^{MN} = \frac{1}{4} \big[ \Gamma^M , \Gamma^N \big]$ is
 the Lorentz generator in ten dimensions 
 and $\omega_{\mathcal{M}, MN}$ is the spin connection.
The non-zero components of the spin connection are 
 \begin{align}
  \omega_{a , mn} &= - \Omega_{mn a} , &
  \omega_{m , ab} &= \frac{1}{2} x^n O_{nmba} , \notag \\
  \omega_{a , mb} &= \frac{1}{2} x^n O_{mn ab} , &
  \omega_{a , bc} &= - \frac{1}{2} x^m O_{mn bc} \Omega^n{}_{a} ,
  \label{eq:spin.connection}
 \end{align}
where 
$O_{mnab}\equiv\Omega_m{}^p{}_a \Omega_{pnb} - \Omega_m{}^p{}_b \Omega_{pna}$ 
is the commutator of the $\Omega$-matrices.
The condition $O_{mnab}=0$ implies that the associated $U(1)^6$ vector fields 
$\Omega^m_{a}\partial_m$ commute with each other.

After the dimensional reduction of the Lagrangian~\eqref{eq:10d.Lag} to
 four dimensions, the Wick rotation 
 and adding the theta term, 
we obtain the following deformed Lagrangian:
\begin{align}
 \mathcal{L}_{\Omega} = \frac{1}{\kappa} \mathrm{Tr} \Big[ &\, \frac{1}{4} F^{mn} F_{mn} + \frac{i \theta g^2}{32 \pi^2} F^{mn} \widetilde{F}_{mn} + \Lambda^A \sigma^{m} D_{m} \bar{\Lambda}_A + \frac{1}{2} \big( D_{m} \varphi_a - g F_{mn} \Omega^{n}_a \big)^2 \notag \\
 &\, - \frac{g}{2} (\Sigma_a )^{AB} \bar{\Lambda}_A [ \varphi_a , \bar{\Lambda}_B ] - \frac{g}{2} (\bar{\Sigma}_a )_{AB} \Lambda^A [ \varphi_a , \Lambda^B ] \notag \\
 &\, - \frac{g^2}{4} \Big( [ \varphi_a , \varphi_b ] + i \Omega^{m}_a D_{m} \varphi_b - i \Omega^{m}_b D_{m} \varphi_a - ig F_{mn} \Omega^{m}_a \Omega^{n}_b \Big)^2 \notag \\
 &\, - \frac{ig}{2} \Omega^{m}_a \big( ( \Sigma_a )^{AB} \bar{\Lambda}_A D_{m} \bar{\Lambda}_B + (\bar{\Sigma}_a )_{AB} \Lambda^A D_{m} \Lambda^B \big) \notag \\
 &\, + \frac{ig}{4} \Omega_{mn a} \big( (\Sigma_a )^{AB} \bar{\Lambda}_A \bar{\sigma}^{mn} \bar{\Lambda}_B + (\bar{\Sigma}_a )_{AB} \Lambda^A \sigma^{mn} \Lambda^B \big) \Big] \label{eq:def.Lag1}
  + \mathcal{L}_O 
 ,
\end{align}
where 
\begin{align}
 \mathcal{L}_O = \frac{1}{\kappa} \mathrm{Tr} \Big[ 
 & - \frac{ig^3}{8} \Big\{ (\bar{\Sigma}^{a} \Sigma^{bc})_{AB} \Lambda^{\alpha A} \Lambda_{\alpha}^B 
     + (\Sigma^a \bar{\Sigma}^{bc})^{AB} \bar{\Lambda}_{\alpha A} \bar{\Lambda}^{\dot{\alpha}}_B \Big\} x^p \Omega^n_a O_{pnbc} \notag \\
 & + \frac{ig^3}{16} \Big\{ (\bar{\Sigma}^a)_{AB} (\sigma^m \bar{\sigma}^n)_{\alpha}{}^{\beta} \Lambda^{\alpha A} \Lambda_{\beta}^B 
     + (\Sigma^a )^{AB} (\bar{\sigma}^m \sigma^n )^{\dot{\alpha}}{}_{\dot{\beta}} \bar{\Lambda}_{\dot{\alpha} A} \bar{\Lambda}^{\dot{\beta}}_B \Big\}
     x^p \Omega_m^b O_{pnab} \notag \\
 & - \frac{ig^4}{8} \Big\{ (\Sigma^{ab})^A{}_B \bar{\sigma}^{m \dot{\alpha}\alpha} \bar{\Lambda}_{\dot{\alpha} A} \Lambda_{\alpha}^B 
     + (\bar{\Sigma}^{ab})_A{}^B \sigma^m_{\alpha \dot{\alpha}} \Lambda^{\alpha A} \bar{\Lambda}^{\dot{\alpha}}_B \Big\} 
     x^p \Omega_m^c \Omega^n_c O_{pnab} \Big].
\end{align}
The terms in $\mathcal{L}_O$ arise from the part of the spin 
connection~\eqref{eq:spin.connection} and are proportional to the 
commutator $O_{mnab}$.

The deformed theory has no supersymmetry in general 
since the $\Omega$-background breaks the Poincar\'{e} symmetry. 
However, a part of the supersymmetry can be recovered 
by choosing the parameters of the background.
For example, when the $\Omega$-matrices are self-dual 
and satisfy the commuting conditions $O_{mnab}=0$, 
the deformed theory reduces to the one obtained in \cite{Ito:2009ac}.
We can show that it preserves a half of $\mathcal{N}=4$ supersymmetry, 
which is given by
\begin{align}
 \delta A_m &= - \bar{\xi}_A \bar{\sigma}_m \Lambda^A \notag , \\
 \delta \Lambda^A &= i (\Sigma_a )^{AB} \sigma^m \bar{\xi}_B \big( D_m \varphi_a - g F_{mn} \Omega^n_a \big) \notag , \\
 \delta \bar{\Lambda}_A &= \bar{\sigma}^{mn} \bar{\xi}_A F_{mn} + ig (\bar{\Sigma}_{ab})_A{}^B \bar{\xi}_B
  \big( \big[ \varphi_a , \varphi_b \big] + i \Omega^m_a D_m \varphi_b - i \Omega^m_b D_m \varphi_a - ig \Omega^m_a \Omega^n_b F_{mn} \big) \notag , \\
 \delta \varphi_a &= - i \bar{\xi}_A (\Sigma_a )^{AB} \bar{\Lambda}_B - g \bar{\xi}_A \bar{\sigma}_m \Lambda^A \Omega^m_a . \label{eq:OmegaSD.SUSY}
\end{align}

In the case of the non-(anti-)self-dual $\Omega$-matrices,
the theory has supersymmetry by introducing 
the R-symmetry Wilson line gauge field 
and by choosing the deformation parameters, 
as in the case of the $\mathcal{N}=2$ theory~\cite{Nekrasov:2003rj,Ito:2010vx}.
The R-symmetry Wilson line is introduced by 
gauging the subgroup $SO(6)$ of the ten-dimensional Lorentz group 
with a constant gauge field $(\mathcal{A}_a)^A{}_B$, 
which takes values in the adjoint representation of $SU(4)_I \sim SO(6)$.
In the ten-dimensional theory, 
the Wilson line modifies the gauge covariant derivative 
along the $x^{a+4}$ directions.
For the spinor fields the gauge covariant derivative is changed as 
\begin{align}
 D_{a+4} \rightarrow D_{a+4} + i \mathcal{A}_{a} .
\end{align}
The gauge covariant derivative for the gauge field is also modified as
\begin{align}
 D_{a+4} \rightarrow D_{a+4} + i \mathcal{A}^{\text{vec}}_{a} , 
 \label{eq:WL.scalar}
\end{align}
%{\bf [$D_a\rightarrow D_{a+4}$?]}
where $\mathcal{A}^{\text{vec}}_a$ is the Wilson line gauge field 
in the vector representation of the $SO(6)$,
which is equivalent to the anti-symmetric representation of $SU(4)_I$.
In the viewpoint of the four-dimensional theory, 
the Wilson line shifts the commutator containing the scalar fields 
$\varphi_a$.
For the spinor fields,
it changes the following terms in the Lagrangian \eqref{eq:def.Lag1} as
\begin{align}
 \big[ \varphi_a , \Lambda^{A} \big] & \rightarrow \big[ \varphi_a , \Lambda^{A} \big] + (\mathcal{A}_a )^{A}{}_{B} \Lambda^{B} \notag , \\
 \big[ \varphi_a , \bar{\Lambda}_{A} \big] & \rightarrow \big[ \varphi_a , \bar{\Lambda}_{A} \big] - \bar{\Lambda}_{B} (\mathcal{A}_a )^{B}{}_{A} .
\end{align}
In contrast to the $\mathcal{N}=2$ case, 
the commutator of the scalar fields $\big[ \varphi_a , \varphi_b \big]$ is 
also changed due to \eqref{eq:WL.scalar}. 
Its shift is given by 
\begin{align}
 \big[ \varphi_a , \varphi_b \big] & \rightarrow \big[ \varphi_a , \varphi_b \big] - \frac{1}{2} \left( (\Sigma_b \bar{\Sigma}_c )^{A}{}_{B} \varphi_{c} (\mathcal{A}_a )^{B}{}_{A} - (\Sigma_a \bar{\Sigma}_c )^{A}{}_{B} \varphi_{c} (\mathcal{A}_b )^{B}{}_{A} \right) .
\end{align} 
Finally, we obtain the Lagrangian in the $\Omega$-background 
with the R-symmetry Wilson line as
\begin{align}
 \mathcal{L}_{(\Omega,\mathcal{A})} = \frac{1}{\kappa} \mathrm{Tr} \Big[ 
 &\, \frac{1}{4} F^{mn} F_{mn} + \frac{i \theta g^2}{32 \pi^2} F^{mn} \widetilde{F}_{mn} + \Lambda^A \sigma^{m} D_{m} \bar{\Lambda}_A + \frac{1}{2} \big( D_{m} \varphi_a - g F_{mn} \Omega^{n}_a \big)^2 \notag \\
 &\, - \frac{g}{2} (\Sigma_a )^{AB} \bar{\Lambda}_A [ \varphi_a , \bar{\Lambda}_B ] - \frac{g}{2} (\bar{\Sigma}_a )_{AB} \Lambda^A [ \varphi_a , \Lambda^B ] \notag \\
 &\, - \frac{g^2}{4} \Big( [ \varphi_a , \varphi_b ] + i \Omega^{m}_a D_{m} \varphi_b - i \Omega^{m}_b D_{m} \varphi_a - ig F_{mn} \Omega^{m}_a \Omega^{n}_b \notag \\
 &\, \qquad \qquad - \frac{1}{2} \big( (\Sigma_b \bar{\Sigma}_c )^{A}{}_{B} \varphi_{c} (\mathcal{A}_a )^{B}{}_{A} - (\Sigma_a \bar{\Sigma}_c )^{A}{}_{B} \varphi_{c} (\mathcal{A}_b )^{B}{}_{A} \big) \Big)^2 \notag \\
 &\, - \frac{ig}{2} \Omega^{m}_a \big( ( \Sigma_a )^{AB} \bar{\Lambda}_A D_{m} \bar{\Lambda}_B + (\bar{\Sigma}_a )_{AB} \Lambda^A D_{m} \Lambda^B \big) \notag \\
 &\, + \frac{ig}{4} \Omega_{mn a} \big( (\Sigma_a )^{AB} \bar{\Lambda}_A \bar{\sigma}^{mn} \bar{\Lambda}_B + (\bar{\Sigma}_a )_{AB} \Lambda^A \sigma^{mn} \Lambda^B \big) \notag \\
 &\, + \frac{g}{2} (\Sigma_a )^{AB} \bar{\Lambda}_A \bar{\Lambda}_{D} (\mathcal{A}_a )^{D}{}_{B} - \frac{g}{2} (\bar{\Sigma}_a )_{AB} \Lambda^A (\mathcal{A}_a )^{B}{}_{D} \Lambda^{D} \Big] 
  + \mathcal{L}_O 
 . \label{eq:def.Lag2}
\end{align}

Note that one can recover the $\Omega$-deformed $\mathcal{N}=2$ 
super Yang-Mills theory~\cite{Nekrasov:2003rj,Ito:2010vx}
from the Lagrangian \eqref{eq:def.Lag2} by 
the $\mathbb{Z}_2$ orbifold projection~\cite{Billo:2006jm,Ito:2010vx}. 
To see this, we decompose the $\mathcal{N}=4$ vector multiplet 
$\big( A_m , \Lambda^A , \bar{\Lambda}_A , \varphi_a \big)$ into 
the $\mathcal{N}=2$ vector multiplet 
and the $\mathcal{N}=2$ adjoint hypermultiplet.
We consider the subgroup 
$SU(2)_I \times SU(2)_{I'}$ of $SU(4)_I$
such that the $SU(4)_I$ index $A=1,2$ corresponds to that of $SU(2)_I$ and 
$A=3,4$ to that of $SU(2)_{I'}$.
We label the indices of the fundamental representations of 
$SU(2)_I$ and $SU(2)_{I'}$ as $A'$ and $\hat{A}$ respectively.
We define $\varphi^{AB} ( = -\varphi^{BA})$ and 
$\bar{\varphi}_{AB} ( = -\bar{\varphi}_{BA})$ by
\begin{align}
 \varphi^{AB} 
 & = 
 \frac{i}{\sqrt{2}} (\Sigma_a )^{AB} \varphi_a , 
 &
 \quad \bar{\varphi}_{AB} 
 & = 
 - \frac{i}{\sqrt{2}} (\bar{\Sigma}_a)_{AB} \varphi_a ,
 \label{eq:twist.scalar}
\end{align}
and decompose them as 
\begin{align}
 \varphi^{AB} &= \begin{pmatrix}
  \varphi \epsilon^{A'B'} & \varphi^{A'\hat{B}} \\
  \varphi^{\hat{A}B'} & - \bar{\varphi} \epsilon^{\hat{A}\hat{B}} 
  \end{pmatrix} , &
 \bar{\varphi}_{AB} &= \begin{pmatrix}
  \bar{\varphi} \epsilon_{A'B'} & \bar{\varphi}_{A'\hat{B}} \\
  \bar{\varphi}_{\hat{A}B'} & - \varphi \epsilon_{\hat{A}\hat{B}} 
  \end{pmatrix} .
 \label{eq:twist.scalar2}
\end{align}
We note that 
$\varphi^{A'\hat{B}}$ and $\bar{\varphi}_{\hat{A}B'}$ are related by 
\begin{align}
 \varphi^{A'\hat{B}} = \epsilon^{A'C'} \epsilon^{\hat{B}\hat{D}}
 \bar{\varphi}_{\hat{D}C'} , 
\end{align}
which is shown by using \eqref{eq:Sigma.relation}. 
Under this decomposition, the $\mathcal{N}=4$ vector multiplet is divided 
into the ${\cal N}=2$ vector multiplet 
$(A_m , \Lambda^{A'} , \bar{\Lambda}_{A'} , \varphi , \bar{\varphi})$ 
and the ${\cal N}=2$ adjoint hypermultiplet 
$(\Lambda^{\hat{A}} , \bar{\Lambda}_{\hat{A}} , \varphi^{\hat{A}B'})$.
We also define $\Omega_{mn}, \bar{\Omega}_{mn}, \Omega^{A'\hat{B}}_{mn}$ and 
$(\mathcal{A})^C{}_D, (\bar{\mathcal{A}})^C{}_D,
(\mathcal{A}^{A'\hat{B}})^C {}_D$ from $\Omega_{mna}$ and 
$(\mathcal{A}_a)^C{}_D$ respectively. 
Now we consider 
the $\mathbb{Z}_2$ subgroup of $SU(2)_{I'}$, 
which changes the signs of the fields and parameters 
with odd $\mathbb{Z}_2$ charges. 
Under the $\mathbb{Z}_2$ projection, 
the hypermultiplet, $\Omega^{A'\hat{B}}_{mn}$, the off-diagonal blocks
$({\cal A})^{A'}{}_{\hat{B}}$, $({\cal A})^{\hat{A}}{}_{B'}$, 
$(\bar{\cal A})^{A'}{}_{\hat{B}}$, $(\bar{\cal A})^{\hat{A}}{}_{B'}$ 
of ${\cal A}$, $\bar{\cal A}$ and the diagonal blocks 
$(\mathcal{A}^{A'\hat{B}})^{C'}{}_{D'}$, 
$(\mathcal{A}^{A'\hat{B}})^{\hat{C}}{}_{\hat{D}}$ 
of 
$\mathcal{A}^{A'\hat{B}}$ are projected out.
After the projection 
and imposing the commuting condition for the $\Omega$-matrices as 
\begin{align}
 \Omega_{mn} \bar{\Omega}_{np} - \bar{\Omega}_{mn} \Omega_{np} = 0 , 
 \label{eq:N2.Omega} 
\end{align}
we get the Lagrangian of the $\Omega$-deformed 
$\mathcal{N}=2$ super Yang-Mills theory obtained 
in~\cite{Nekrasov:2003rj,Ito:2010vx}.
The $\Omega$-matrices satisfying \eqref{eq:N2.Omega} 
are expressed by the $\epsilon$-parameters $\epsilon_1, \epsilon_2$ as 
\begin{align}
 \Omega_{mn} = \frac{1}{2\sqrt{2}} \begin{pmatrix}
  0 & i \epsilon_1 & 0 & 0 \\
  -i \epsilon_1 & 0 & 0 & 0 \\
  0 & 0 & 0 & -i \epsilon_2 \\
  0 & 0 & i \epsilon_2 & 0 \end{pmatrix}
  , \quad 
 \bar{\Omega}_{mn} = \frac{1}{2\sqrt{2}} \begin{pmatrix}
  0 & -i \bar{\epsilon}_1 & 0 & 0 \\
  i \bar{\epsilon}_1 & 0 & 0 & 0 \\
  0 & 0 & 0 & i \bar{\epsilon}_2 \\
  0 & 0 & -i \bar{\epsilon}_2 & 0 \end{pmatrix} .
 \label{eq:Omega.epsilon}
\end{align}
The $\Omega$-deformed $\mathcal{N}=2$ super Yang-Mills theory has one 
 fermionic charge
 when we choose the R-symmetry Wilson line gauge field as~\cite{LoMaNe,Nekrasov:2003rj,Ito:2010vx}
\begin{align}
 & \mathcal{A}^{A'}{}_{B'} = - \frac{1}{2} \Omega_{mn} (\bar{\sigma}^{mn})^{A'}{}_{B'} , & 
 & \bar{\mathcal{A}}^{A'}{}_{B'} = - \frac{1}{2} \bar{\Omega}_{mn} (\bar{\sigma}^{mn})^{A'}{}_{B'} , 
 \label{eq:N2.WL} 
\end{align}
and $(\mathcal{A}^{A'\hat{B}})^{C'}{}_{\hat{D}}=(
\mathcal{A}^{A'\hat{B}})^{\hat{C}}{}_{D'}=0$.
We note that the other components of the Wilson line gauge field 
do not appear in the Lagrangian 
because they couple only with the hypermultiplet.

One can also obtain 
the mass deformed ${\cal N}=4$ theory or the $\mathcal{N}=2^*$ theory 
in the $\Omega$-background when the $\mathbb{Z}_2$-projected 
$\Omega$-background satisfies \eqref{eq:N2.Omega} and 
the R-symmetry Wilson line gauge fields take the following form: 
\begin{align} 
 \mathcal{A}^{A}{}_{B} &= 
  \begin{pmatrix}
   \mathcal{A}^{A'}{}_{B'} & 0 \\
   0 & M^{\hat{A}}{}_{\hat{B}}
  \end{pmatrix} , &
 \bar{\mathcal{A}}^{A}{}_{B} &= 
  \begin{pmatrix}
   \bar{\mathcal{A}}^{A'}{}_{B'} & 0 \\
   0 & \bar{M}^{\hat{A}}{}_{\hat{B}}
  \end{pmatrix} , &
 (\mathcal{A}^{C'\hat{D}})^{A}{}_{B} &= 0 .
  \label{eq:Wilson.cond}
\end{align}
Here 
$\mathcal{A}^{A'}{}_{B'}$ and $\bar{\mathcal{A}}^{A'}{}_{B'}$ 
are the $\mathcal{N}=2$ R-symmetry Wilson lines \eqref{eq:N2.WL}. 
The mass deformation parameters of the $\mathcal{N}=2^*$ theory are given by 
$M^{\hat{A}}{}_{\hat{B}}$ and $\bar{M}^{\hat{A}}{}_{\hat{B}}$. 
Taking them 
to be proportional to $\tau^3$ by the $SU(2)_{I'}$ transformation as 
\begin{align}
 M^{\hat{A}}{}_{\hat{B}} &= \begin{pmatrix} m & 0 \\ 0 & -m \end{pmatrix} , & 
 \bar{M}^{\hat{A}}{}_{\hat{B}} &= \begin{pmatrix} \bar{m} & 0 \\ 0 & -\bar{m} \end{pmatrix} ,  
 \label{mass}
\end{align}
the mass of the hypermultiplet is $\sqrt{m\bar{m}}$. 
The action of the $\Omega$-deformed $\mathcal{N}=2^*$ theory is
invariant under the following supersymmetry transformations: 
%{\bf [transformations?]}:
\begin{align}
 \delta A_{m} &= 
  \bar{\xi} \sigma_{m \alpha A'} \Lambda^{\alpha A'} , \notag \\
 \delta \Lambda_{\alpha}^{A'} &= 
  - \sqrt{2} \bar{\xi} \epsilon^{A'B'} \sigma^{m}_{\alpha B'} \big( D_{m} \varphi - g F_{mn} \Omega^{n} \big) , \notag \\
 \delta \Lambda_{\alpha}^{\hat{A}} &= 
  - \sqrt{2} \bar{\xi} \sigma^{m}_{\alpha B'} D_{m} \varphi^{\hat{A} B'} , \notag \displaybreak[4]\\
 \delta \bar{\Lambda}^{\dot{\alpha}}_{A'} &= 
  - \bar{\xi} (\bar{\sigma}^{mn})^{\dot{\alpha}}{}_{A'} F_{mn} 
  + {ig} \bar{\xi} \delta^{\dot{\alpha}}_{A'} 
    \left( [ \varphi , \bar{\varphi} ] + i \Omega^{m} D_{m} \bar{\varphi} - i \bar{\Omega}^{m} D_{m} \varphi 
    - ig F_{mn} \Omega^{m} \bar{\Omega}^{n} \right)  \notag \\
  & \qquad + ig \bar{\xi} \delta^{\dot{\alpha}}_{B'} [ \varphi^{B' \hat{C}} , \bar{\varphi}_{\hat{C}A'} ] , \notag \\
 \delta \bar{\Lambda}^{\dot{\alpha}}_{\hat{A}} &= 
  - 2ig \bar{\xi} \epsilon^{\dot{\alpha}B'} \left( \big[ \varphi , \bar{\varphi}_{\hat{A} B'} \big] + i \Omega^{m} D_{m} \bar{\varphi}_{\hat{A} B'} 
    - \mathcal{A}^{D'}{}_{B'} \bar{\varphi}_{\hat{A}D'} - M^{\hat{D}}{}_{\hat{A}} \bar{\varphi}_{\hat{D}B'} \right) , \notag \\
 \delta \varphi &= 
  g \bar{\xi} \sigma^{m}_{\alpha A'} \Lambda^{\alpha A'} \Omega_{m} , \notag \\
 \delta \bar{\varphi} &= 
  - \sqrt{2} \bar{\xi} \epsilon^{\dot{\alpha}A'} \bar{\Lambda}_{\dot{\alpha} A'} 
  + g \bar{\xi} \sigma^{m}_{\alpha A'} \Lambda^{\alpha A'} \bar{\Omega}_{m} , \notag \\
 \delta \bar{\varphi}_{A'\hat{B}} &= 
  \sqrt{2} \bar{\xi} \delta_{A'}^{\dot{\alpha}} \bar{\Lambda}_{\dot{\alpha} \hat{B}} .
 \label{eq:def2*.SUSY}
\end{align}
Here 
 $\Omega_m = \Omega_{mn} x^n$ and 
 $\bar{\Omega}_m = \bar{\Omega}_{mn} x^n$. 
The parameter $\bar{\xi}$ 
 is obtained by 
 taking the diagonal part of the 
 $\mathcal{N}=2$ supersymmetry transformation parameters 
 $\bar{\xi}^{\dot{\alpha}}_{A'}$ as 
$\bar{\xi} = \delta_{\dot{\alpha}}^{A'} \bar{\xi}^{\dot{\alpha}}_{A'}$
\cite{Witten:1988ze}.
The transformations \eqref{eq:def2*.SUSY} can be regarded 
as the $\Omega$-deformation of the topologically twisted supersymmetry.

\subsection{Instanton calculus in deformed 
$\mathcal{N}=4$ theory}
We now study the instanton calculus in the deformed $\mathcal{N}=4$ 
$U(N)$ super Yang-Mills theory.
We are interested in the integration over the zero modes around
the instanton solutions in the path integral. 
We first examine the solutions to the equations of motion
around the instantons. 
In the Coulomb branch of the theory, the adjoint scalar fields $\varphi_{a}$ 
have vacuum expectation values (VEVs) 
$\langle\varphi_{a}\rangle=\phi^{0}_{a}$. 
We take the VEVs to be diagonal such that they commute 
with each other. 
It is very difficult to 
solve the equations exactly even in the case where all 
the deformation parameters vanish. 
Instead we expand the fields 
in the coupling constant $g$. 
Its approximation is valid when the VEVs are large 
\cite{Affleck:1980mp}. 
The leading term 
of the gauge field satisfies 
the (anti-)self-dual equation. 
Here we consider the case of the self-dual instanton solution. 
The $g$-expansions of the fields are given by 
\begin{align}
A_m &= g^{-1} A^{(0)}_m + g^1 A^{(1)}_m + \cdots, 
&
\Lambda^A &= g^{-1/2} \Lambda^{(0) A} + g^{3/2} \Lambda^{(1) A} + \cdots, 
\notag\\
\varphi_a &= g^0 \varphi^{(0)}_a + g^{2} \varphi^{(1)}_a + \cdots, 
&
\bar{\Lambda}_A &= g^{1/2} \bar{\Lambda}^{(0)}_A 
+ g^{5/2} \bar{\Lambda}^{(1)}_A + \cdots.
\label{expansion}
\end{align}
Here $A^{(0)}_{m}$ satisfies the self-dual equation: 
\begin{gather}
F^{(0)}_{mn} = \widetilde{F}^{(0)}_{mn}, 
\label{eom1}
\end{gather}
where $F^{(0)}_{mn}$ is the field strength of $A^{(0)}_{m}$. 
{}From the Lagrangian \eqref{eq:def.Lag2} 
the equations of motion for the other fields 
at the leading order in $g$ are obtained as 
\begin{align}
 & \bar{\sigma}^{m \dot{\alpha} \alpha} \nabla_m \Lambda^{(0)}{}_{\alpha}^{A} = 0 , \label{eom2}\\
 & \nabla^2 \varphi^{(0)}_a + F^{(0)}_{mn} \Omega^{mn}_a + (\bar{\Sigma}_a)_{AB} \Lambda^{(0) \alpha A} \Lambda^{(0)}{}_{\alpha}^B = 0 , \label{eom3}\\
 & \sigma^{m}_{\alpha \dot{\alpha}} \nabla_m \bar{\Lambda}^{(0)}{}^{\dot{\alpha}}_A - (\bar{\Sigma}_a )_{AB} \big[ \varphi^{(0)}_a , \Lambda^{(0)}{}_{\alpha}^B \big] \notag\\
 & \hspace{1cm} - i \Omega^m_a (\bar{\Sigma}_a)_{AB} \nabla_m \Lambda^{(0)}{}_{\alpha}^B + \frac{i}{2} \Omega_{mna} (\bar{\Sigma}_a)_{AB} (\sigma^{mn})_{\alpha}{}^{\beta} \Lambda^{(0)}{}_{\beta}^B - (\bar{\Sigma}_a)_{AB} (\mathcal{A}_a)^B{}_C \Lambda^{(0)}{}_{\alpha}^C = 0 ,
\label{eom4}
\end{align}
where 
$\nabla_{m}\ast =\partial_{m}\ast+i[A_{m}^{(0)},\ast]$
is the gauge covariant derivative in the instanton background \eqref{eom1}. 
The parameters $\Omega_{mna}$ and $\mathcal{A}_{a}$ 
are of order $g^0$. 
The equations \eqref{eom1} and \eqref{eom2} are not deformed, 
while only the self-dual part of $\Omega_{mna}$ contributes in the equation 
\eqref{eom3} 
since it is contracted with the self-dual $F^{(0)}_{mn}$. 
Then the solutions to \eqref{eom1}--\eqref{eom3} 
are the same as those in the self-dual $\Omega$-background 
and without the R-symmetry Wilson line, which have been obtained 
in \cite{Ito:2009ac}. 
The solution to \eqref{eom4} 
does not contribute to the instanton effective action 
as we will see later. 

The equation \eqref{eom1} for the instantons is solved by 
the ADHM construction \cite{AtHiDrMa}. 
The %instanton{\bf [remove instanton?]} 
solution with the instanton number $k$ 
is parametrized by 
the position moduli $a'_{m}$ and the size moduli 
$w_{\dot{\alpha}}$, $\bar{w}^{\dot{\alpha}}$. 
Here $a'_{m}$ are the $k\times k$ Hermitian matrices and 
$w_{\dot{\alpha}}$, $\bar{w}^{\dot{\alpha}}$ are 
the $N\times k$ and $k\times N$ complex matrices respectively, 
which are Hermitian conjugate to each other. 
The variables $a'_{m}$, $w_{\dot{\alpha}}$ and $\bar{w}^{\dot{\alpha}}$ 
are called the bosonic ADHM moduli and 
satisfy the bosonic ADHM constraints 
\begin{equation}
(\tau^{\tilde{c}})^{\dot{\alpha}}_{~\dot{\beta}}
(\bar{w}^{\dot{\beta}}w_{\dot{\alpha}}
+\bar{a}^{\prime\dot{\beta}\alpha}a'_{\alpha\dot{\alpha}})=0, 
\quad 
\tilde{c}=1,2,3, 
\label{ADHM}
\end{equation}
where $a'_{\alpha\dot{\alpha}}$ are defined by 
$a'_{\alpha\dot{\alpha}}=\sigma^{m}_{\alpha\dot{\alpha}}a'_{m}$. 
The instanton moduli space 
is described by the variables 
satisfying \eqref{ADHM}, 
divided by the $U(k)$ action 
\begin{gather}
a'_{m}\to u^{-1}a'_{m}\,u,\quad
w_{\dot{\alpha}}\to w_{\dot{\alpha}}\,u,\quad
\bar{w}^{\dot{\alpha}}\to u^{-1}\bar{w}^{\dot{\alpha}},\quad
u\in U(k).
\label{Uk}
\end{gather}
We can also solve the fermionic zero mode equation \eqref{eom2} 
by the ADHM construction \cite{DoHoKhMa}. 
The fermionic zero modes 
are parametrized by the Grassmann-odd matrices 
$\mathcal{M}'{}^{A}_{\alpha}$, $\mu^{A}$ and $\bar{\mu}^{A}$ 
called the fermionic ADHM moduli. 
They are the superpartners of 
$a'_{m}$, $w_{\dot{\alpha}}$ and $\bar{w}^{\dot{\alpha}}$ 
respectively and then have the same size of matrices with their partners. 
The fermionic ADHM moduli satisfy the fermionic ADHM constraints 
\begin{equation}
\bar{\mu}^{A}w_{\dot{\alpha}}+\bar{w}_{\dot{\alpha}}\mu^{A}
+[\mathcal{M}^{\prime\alpha A},a'_{\alpha\dot{\alpha}}]=0.
\label{fADHM}
\end{equation}
The $U(k)$ group also acts on the fermionic moduli as 
\begin{gather}
\mathcal{M}'{}^{A}_{\alpha}\to u^{-1}\mathcal{M}'{}^{A}_{\alpha}\,u,\quad
\mu^{A}\to \mu^{A}\,u,\quad
\bar{\mu}^{A}\to u^{-1}\bar{\mu}^{A}\,. 
\label{Uk2}
\end{gather}
The solution to the equation \eqref{eom3} for the scalar fields 
does not have zero modes 
and is expressed in terms of the ADHM moduli, the scalar VEVs and 
the deformation parameter $\Omega_{mna}$ \cite{Ito:2009ac}. 

By substituting the expansions \eqref{expansion} into \eqref{eq:def.Lag2}, 
the spacetime action $S=\int d^{4}x\,\mathcal{L}_{(\Omega,\mathcal{A})}$ 
is expanded as 
\begin{gather}
S=\biggl(\frac{8\pi^2}{g^2}-i\theta\biggr)k+g^{0}S^{(0)}
+\mathcal{O}(g^2), 
\label{expansion2}
\end{gather}
where $S^{(0)}$ is given by 
\begin{align}
 S^{(0)} = \frac{1}{\kappa} \int \mathrm{d}^{4}x \, \mathrm{Tr} \bigg[ & \frac{1}{2} \big( \nabla_m \varphi^{(0)}_a - F^{(0)}_{mn} \Omega^n_a \big)^2 - \frac{1}{2} (\bar{\Sigma}_a)_{AB} \Lambda^{(0)}{}^{\alpha A} \big[ \varphi_a , \Lambda^{(0)}{}_{\alpha}^{B} \big] \notag \\
 & - \frac{i}{2} \Omega^m_a (\bar{\Sigma}_a)_{AB} \Lambda^{(0)}{}^{\alpha A} \nabla_m \Lambda^{(0)}{}_{\alpha}^B + \frac{i}{4} \Omega_{mna} (\bar{\Sigma}_a)_{AB} \Lambda^{(0)}{}^{\alpha A} (\sigma^{mn})_{\alpha}{}^{\beta} \Lambda^{(0)}{}_{\beta}^B \notag \\
 & - \frac{1}{2} (\bar{\Sigma}_a)_{AB} \Lambda^{(0)}{}^{\alpha A} (\mathcal{A}_a)^B{}_C \Lambda^{(0)}{}_{\alpha}^C \bigg]. 
\label{Szero}
\end{align}
As we mentioned before, 
$S^{(0)}$ does not depend on $\bar{\Lambda}_{A}^{(0)}$. 
We evaluate 
the integral in \eqref{Szero} by 
substituting the solutions to \eqref{eom1}--\eqref{eom3} 
and we express $S^{(0)}$ as the function of the ADHM moduli. 
In the case of the self-dual $\Omega_{mna}$ and 
the vanishing R-symmetry Wilson line, 
\eqref{Szero} is reduced to the one in \cite{Ito:2009ac}. 
In the general case we 
have the contributions from the anti-self-dual part of $\Omega_{mna}$ 
and the Wilson line, which can be computed in a similar way 
as in the case of the ${\cal N}=2$ theory~\cite{Ito:2010vx}. 
Then $S^{(0)}$ is evaluated as 
\begin{align}
S^{(0)}
&= 
\frac{2\pi^2}{\kappa}\mathrm{tr}_k
\biggl[-\biggl(\frac{1}{4} (\bar{\Sigma}_{a})_{AB}
\Bigl(\bar{\mu}^{A}\mu^{B}
+\mathcal{M}^{\prime\alpha A}\mathcal{M}^{\prime B}_{\alpha}\Bigr)
+\bar{w}^{\dot{\alpha}}\phi^{0}_{a} w_{\dot{\alpha}}
-i\Omega^{+mna}[a'_{m},a'_{n}]\biggr)
\notag\\
&\qquad\qquad\quad{}
\times\boldsymbol{L}^{-1}\biggl(\frac{1}{4} (\bar{\Sigma}_{a})_{CD}
\Bigl(\bar{\mu}^{C}\mu^{D}
+\mathcal{M}^{\prime\beta C}\mathcal{M}^{\prime D}_{\beta}\Bigr)
+\bar{w}^{\dot{\beta}}\phi^{0}_{a} w_{\dot{\beta}}
-i\Omega^{+pqa}[a'_{p},a'_{q}]\biggr)
\notag\\
&\qquad\qquad\quad{}
+\frac{i}{8} (\bar{\Sigma}^a)_{AB} \Omega_{mna}
(\sigma^{mn})_{\alpha}{}^{\beta}
\mathcal{M}^{\prime\alpha}{}^{A}\mathcal{M}^{\prime B}_{\beta}
+\frac{1}{4}\Omega^{-mna}\Omega^{-}_{mna}
\bar{w}^{\dot{\alpha}}w_{\dot{\alpha}}
\notag\\
&\qquad\qquad\quad{}
+\frac{1}{2} (\bar{\Sigma}^a)_{AB}
\bar{\mu}^A \phi^0_a \mu^B
- \bar{w}_{\dot{\alpha}} \phi^0_a \phi^0_a w^{\dot{\alpha}}
+i\Omega^{-}_{mna}(\bar{\sigma}^{mn})^{\dot{\alpha}}{}_{\dot{\beta}}
\bar{w}^{\dot{\beta}}\phi^{0}_{a}w_{\dot{\alpha}}
\notag\\
&\qquad\qquad\quad{}
+\Omega^{mna}\Omega_{mpa}a'_{n}a^{\prime p}
-m_{AB}\Bigl(2\bar{\mu}^{A}\mu^{B}
+\mathcal{M}^{\prime \alpha A}\mathcal{M}^{\prime B}_{\alpha}\Bigr)
\biggr],
\label{Szero2}
\end{align}
where $\mathrm{tr}_k$ denotes the trace for the $k\times k$ matrix. 
We use the same normalization factor $\kappa$ for 
the $U(k)$ generators. 
$\Omega^{\pm}_{mna}=\frac{1}{2}(\Omega_{mna}\pm\widetilde{\Omega}_{mna})$ 
is the (anti-)self-dual part of $\Omega_{mna}$. 
The operator $\boldsymbol{L}$ acting on the $k\times k$ matrix 
is defined by 
\begin{gather}
\boldsymbol{L}
=
\frac{1}{2}\bigl\{\bar{w}^{\dot{\alpha}}w_{\dot{\alpha}},\ast\bigr\}
+\bigl[a^{\prime m},[a'_{m},\ast]\bigr].
\end{gather}
The parameters $m_{AB}$, which are symmetric in the indices $A$ and $B$,
are expressed by the Wilson line gauge field 
$(\mathcal{A}_{a})^{A}{}_{B}$ as 
%The mass parameter $m_{AB}$ for the fermionic moduli 
%is given by the 
%Wilson line gauge field 
%$(\mathcal{A}_{a})^{A}{}_{B}$ as 
\begin{align}
m_{AB}=\frac{1}{8}\bigl((\bar{\Sigma}_a)_{AC}(\mathcal{A}_a)^{C}{}_{B}
+(\bar{\Sigma}_a)_{BC}(\mathcal{A}_a)^{C}{}_{A}\bigr). 
\label{mAB}
\end{align}
In \eqref{Szero2} the ADHM moduli obey the ADHM constraints. 
We introduce the 
auxiliary variables 
$D^{\tilde{c}}$ 
and $\bar{\psi}^{\dot{\alpha}}_{A}$ as the Lagrange multipliers 
for the constraints \eqref{ADHM} and \eqref{fADHM} respectively 
such that we can regard the ADHM moduli as the independent variables 
in the path integral. 
We also introduce the auxiliary variables $\chi_{a}$ 
in the path integral such that the 
$\boldsymbol{L}^{-1}$-terms in \eqref{Szero2} become the Gaussian form. 
Then $S^{(0)}$ can be rewritten as 
\begin{align}
S^{(0)}_{\text{eff}}&=
\frac{2\pi^2}{\kappa}\mathrm{tr}_{k}\Bigl[ 
-\bigl(\bigl[\chi_{a},a'_{m}\bigr]-i\Omega_{mna}a^{\prime n}\bigr)
\bigl(\bigl[\chi_{a}, a^{\prime m}\bigr]-i\Omega^{mp}_{a}a'_{p}\bigr)
\notag\\
&\qquad\quad{}
-\frac{1}{4}\mathcal{M}^{\prime\alpha A}\Bigl((\bar{\Sigma}_a)_{AB}
\Bigl(\bigl[\chi_{a},\mathcal{M}'{}_{\alpha}^{B}\bigr]
-\frac{i}{2}\Omega_{mna}(\sigma^{mn})_{\alpha}{}^{\beta}
\mathcal{M}'{}_{\beta}^{B}\Bigr)+4m_{AB}\mathcal{M}'{}_{\alpha}^{B}\Bigr)
\notag\\
&\qquad\quad{}
+\biggl(\chi_{a}\bar{w}^{\dot{\alpha}}-\bar{w}^{\dot{\alpha}}\phi^{0}_{a}
-\frac{i}{2}\Omega_{mna}(\bar{\sigma}^{mn})^{\dot{\alpha}}{}_{\dot{\beta}}
\bar{w}^{\dot{\beta}}\biggr)
\biggl(w_{\dot{\alpha}}\chi_{a}-\phi^{0}_{a}w_{\dot{\alpha}}
-\frac{i}{2} \Omega_{pqa}(\bar{\sigma}^{pq})^{\dot{\gamma}}{}_{\dot{\alpha}}
w_{\dot{\gamma}}\biggr)
\notag\\
&\qquad\quad{}
+\frac{1}{2}\bar{\mu}^{A}\Bigl((\bar{\Sigma}_a)_{AB}
\bigl(\mu^{B}\chi_{a}-\phi^{0}_{a}\mu^{B}\bigr)-4m_{AB}\mu^{B}\Bigr)
\notag\\
&\qquad\quad{}
-i\bar{\psi}^{\dot{\alpha}}_{A}
\bigl(\bar{\mu}^{A}w_{\dot{\alpha}}+\bar{w}_{\dot{\alpha}}\mu^{A}
+\bigl[\mathcal{M}'^{\alpha A},a'_{\alpha \dot{\alpha}}\bigr]\bigr)
+iD^{\tilde{c}}(\tau^{\tilde{c}})^{\dot{\alpha}}{}_{\dot{\beta}}
\bigl(\bar{w}^{\dot{\beta}}w_{\dot{\alpha}}
+\bar{a}^{\prime\dot{\beta}\alpha} a'_{\alpha \dot{\alpha}}\bigr)
-i\zeta^{\tilde{c}}D^{\tilde{c}}
\Bigr],
\label{eq:inst.eff}
\end{align}
which is called the instanton effective action. 
Here we have 
introduced the Fayet-Iliopoulos (FI) parameters $\zeta^{\tilde{c}}$ 
to resolve the small instanton singularity, 
which modifies the bosonic ADHM constraints \eqref{ADHM} as 
\begin{gather}
(\tau^{\tilde{c}})^{\dot{\alpha}}_{~\dot{\beta}}
(\bar{w}^{\dot{\beta}}w_{\dot{\alpha}}
+\bar{a}^{\prime\dot{\beta}\alpha}a'_{\alpha\dot{\alpha}})=\zeta^{\tilde{c}}. 
\label{mADHM}
\end{gather}
The parameters 
$\zeta^{\tilde{c}}$ 
are also interpreted as the noncommutativity 
parameters in spacetime~\cite{NeSc}. 
In the instanton effective action (\ref{eq:inst.eff}), the $\Omega$-background
parameters $\Omega_{mna}$ give the mass terms for the bosonic moduli 
$(a'_m, w_{\dot{\alpha}}, \bar{w}^{\dot{\alpha}})$ and 
for the fermionic moduli 
$\mathcal{M}'{}^{A}_{\alpha}$, while the parameters $m_{AB}$ give 
the mass terms for the fermionic moduli 
$(\mathcal{M}'{}^{A}_{\alpha},\mu^A,\bar{\mu}^A)$. 

For the case with the self-dual $\Omega$-background and without the Wilson 
line the effective action (\ref{eq:inst.eff}) is reduced 
to the one obtained in \cite{Ito:2009ac}, where 
the action is invariant under the deformed supersymmetry. 
We can also reduce the effective action \eqref{eq:inst.eff} 
to that in $\mathcal{N}=2$ case by 
the $\mathbb{Z}_{2}$ orbifold projection which acts on 
the parameters with the 
%the indices of the R-symmetry subgroup 
$SU(2)_{I'}$ indices. 
Here the moduli variables and the deformation parameters 
having odd $\mathbb{Z}_{2}$ charges are projected out. 
To describe the $\mathcal{N}=2$ theory, 
we decompose 
the fermionic moduli $\mathcal{M}^{\prime A}_{\alpha}$, $\mu^{A}$, 
the auxiliary variables $\chi_{a}$, $\bar{\psi}^{\dot{\alpha}}_{A}$ 
and the scalar VEVs $\phi^{0}_{a}$ to 
$(\mathcal{M}^{\prime A'}_{\alpha},\mathcal{M}^{\prime\hat{A}}_{\alpha})$, 
$(\mu^{A'},\mu^{\hat{A}})$, $(\chi,\bar{\chi},\chi^{A' \hat{A}})$, 
$(\bar{\psi}^{\dot{\alpha}}_{A'},\bar{\psi}^{\dot{\alpha}}_{\hat{A}})$ 
and $(\phi^{0},\bar{\phi}^{0},\phi^{A' \hat{A}}_{0})$ 
respectively. 
The deformation parameters $\Omega_{mna}$ and 
$(\mathcal{A}_{a})^{A}{}_{B}$ are 
decomposed as in the previous subsection. 
Then the mass parameter $m_{AB}$ is decomposed to 
$(m_{A' B'},m_{A' \hat{A}},m_{\hat{A}\hat{B}})$. 
Here 
$m_{A' B'}$ is related to the $\mathcal{N}=2$ Wilson line gauge field 
$\bar{\mathcal{A}}^{A'}{}_{B'}$ as 
$m_{A' B'}=-2\sqrt{2}i\epsilon_{A' C'}\bar{\mathcal{A}}^{C'}{}_{B'}$. 
Under the orbifold projection the moduli
$(\mathcal{M}^{\prime\hat{A}}_{\alpha},\mu^{\hat{A}}, 
\chi^{A' \hat{A}},\bar{\psi}^{\dot{\alpha}}_{\hat{A}})$, 
the scalar VEVs $\phi^{A' \hat{A}}_{0}$ 
and the parameters $(\Omega^{A' \hat{A}}_{mn}, m_{A' \hat{A}})$ 
are projected out. 
The parameter $m_{\hat{A}\hat{B}}$ survives in the projection 
but does not contribute to the effective action since it only couples to the 
fermionic moduli $\mathcal{M}^{\prime\hat{A}}_{\alpha}$ and $\mu^{\hat{A}}$. 
Then \eqref{eq:inst.eff}
is reduced to the deformed 
$\mathcal{N}=2$ effective action, which has 
one deformed supersymmetry 
under the conditions \eqref{eq:N2.Omega} and \eqref{eq:N2.WL} 
\cite{Nekrasov:2002qd, Ito:2010vx}. 

In the case without the $\Omega$-background and 
with the non-zero Wilson line parameters $M^{\hat{A}}{}_{\hat{B}}$, 
we find that the instanton effective action \eqref{eq:inst.eff} becomes 
that of the undeformed $\mathcal{N}=2^{*}$ theory \cite{Hollowood:2002zv}. 
The $\mathcal{N}=2$ mass parameters $M^{\hat{A}}{}_{\hat{B}}$ 
are related with $m_{AB}$ via 
$M^{\hat{A}}{}_{\hat{B}}=-2\sqrt{2}i
\epsilon^{\hat{A}\hat{C}}m_{\hat{C}\hat{B}}$. 
We note that the effective action 
\eqref{eq:inst.eff} does not depend on $\bar{M}^{\hat{A}}{}_{\hat{B}}$ 
since the terms in the spacetime action 
depending on $\bar{M}^{\hat{A}}{}_{\hat{B}}$ are of higher order 
in the coupling constant expansion \eqref{expansion2}. 
We have to set $\phi^{A' \hat{A}}_{0}=0$ since it breaks the 
$\mathcal{N}=2$ supersymmetry 
due to the mass terms for the scalar fields in the hypermultiplet 
\cite{Hollowood:2002zv}. 

In the $\Omega$-background and the Wilson lines satisfying 
$\Omega^{A' \hat{A}}_{mn}=0$, 
\eqref{eq:N2.Omega}, \eqref{eq:N2.WL} and \eqref{eq:Wilson.cond}, 
we obtain the instanton effective action of 
the deformed $\mathcal{N}=2^{*}$ theory. 
The action \eqref{eq:inst.eff} is invariant under the following
supersymmetry transformations: 
\begin{alignat}{2}
\delta a'_{\alpha\dot{\alpha}}&=
\bar{\xi}\epsilon_{\dot{\alpha}A'}\mathcal{M}_{\alpha}^{\prime A'},
&
\delta\mathcal{M}_{\alpha}^{\prime A'}&=
\sqrt{2}\bar{\xi}\epsilon^{A' \dot{\alpha}}\bigl(
2i[a'_{\alpha\dot{\alpha}},\chi]
\notag\\
&&&\qquad{}
-\Omega^{+}_{mn}(\sigma^{mn})_{\alpha}{}^{\beta}a'_{\beta\dot{\alpha}}
+\Omega^{-}_{mn}(\bar{\sigma}^{mn})^{\dot{\beta}}{}_{\dot{\alpha}}
a'_{\alpha\dot{\beta}}\bigr), 
\notag\\[2mm]
\delta w_{\dot{\alpha}}&=\bar{\xi}\epsilon_{\dot{\alpha}A'}\mu^{A'},
&
\delta\mu^{A'}&=
\sqrt{2}\bar{\xi}\epsilon^{A'\dot{\alpha}}
\bigl(2i(w_{\dot{\alpha}}\chi-\phi^{0}w_{\dot{\alpha}})
+\Omega^{-}_{mn}(\bar{\sigma}^{mn})^{\dot{\beta}}{}_{\dot{\alpha}}
w_{\dot{\beta}}\bigr),
\notag\\[2mm]
\delta\bar{w}^{\dot{\alpha}}&=
\bar{\xi}\delta_{A'}^{\dot{\alpha}}\bar{\mu}^{A'},
&
\delta\bar{\mu}^{A'}&=
-\sqrt{2}\bar{\xi}\delta^{A'}_{\dot{\alpha}}\bigl(
2i(\chi\bar{w}^{\dot{\alpha}}-\bar{w}^{\dot{\alpha}}\phi^{0})
+\Omega^{-}_{mn}(\bar{\sigma}^{mn})^{\dot{\alpha}}{}_{\dot{\beta}}
\bar{w}^{\dot{\beta}}\bigr),
\notag\\[2mm]
\delta\chi&=0, & & 
\notag\\[2mm]
\delta\bar{\chi}&=
-\sqrt{2}\bar{\xi}\delta^{A'}_{\dot{\alpha}}\bar{\psi}_{A'}^{\dot{\alpha}},
&
\delta\bar{\psi}_{A'}^{\dot{\alpha}}&=
\bar{\xi}\bigl(
-i\delta_{A'}^{\dot{\alpha}}[\bar{\chi},\chi]
+(\tau^{\tilde{c}})^{\dot{\alpha}}{}_{A'}D^{\tilde{c}}
+i\delta_{B'}^{\dot{\alpha}}[\chi^{B' \hat{A}}, \bar{\chi}_{\hat{A}A'}]
\bigr),
\notag\\[2mm]
\delta\chi^{\dot{\alpha}\hat{A}}&=
\sqrt{2}\bar{\xi}\bar{\psi}^{\dot{\alpha}\hat{A}},
&
\delta\bar{\psi}^{\dot{\alpha}\hat{A}}&=
\bar{\xi}\bigl(2i[\chi^{\dot{\alpha}\hat{A}},\chi]
-\Omega^{-}_{mn}(\bar{\sigma}^{mn})^{\dot{\alpha}}{}_{\dot{\beta}}
\chi^{\dot{\beta}\hat{A}}-2iM^{\hat{A}}{}_{\hat{B}}\chi^{\dot{\alpha}\hat{B}}
\bigr),
\notag\\[2mm]
\delta D^{\tilde{c}}&=
\frac{i}{\sqrt{2}}\bar{\xi}(\tau^{\tilde{c}})^{A'}{}_{\dot{\alpha}}
[\bar{\psi}_{A'}^{\dot{\alpha}},\chi]
&{}-\sqrt{2}i&\bar{\xi}(\tau^{\tilde{c}})^{A'}{}_{\dot{\alpha}}
[\bar{\psi}^{\dot{\alpha}\hat{A}},\bar{\chi}{}_{\hat{A}A'}]
+\frac{1}{\sqrt{2}}\bar{\xi}\epsilon^{\tilde{c}\tilde{d}\tilde{e}}\Omega^{mn}
\bar{\eta}^{\tilde{d}}_{mn}(\tau^{\tilde{e}})^{A'}{}_{\dot{\alpha}}
\bar{\psi}^{\dot{\alpha}}_{A'}, 
\notag\\[2mm]
\delta\mathcal{M}^{\prime\hat{A}}_{\alpha}&=
-2\sqrt{2}i\bar{\xi}[a'_{\alpha\dot{\alpha}}, \chi^{\dot{\alpha}\hat{A}}],&
\quad 
\delta\mu^{\hat{A}}&=
-2\sqrt{2}i\bar{\xi}w_{\dot{\alpha}}\chi^{\dot{\alpha}\hat{A}},
\quad 
\delta\bar{\mu}^{\hat{A}}=
-2\sqrt{2}i\bar{\xi}\epsilon^{\hat{A}\hat{B}}
\bar{\chi}_{\hat{B}\dot{\alpha}}\bar{w}^{\dot{\alpha}},
\label{defN4BRST}
\end{alignat}
where the anti-self-dual 't Hooft $\eta$-symbol $\bar{\eta}^{\tilde{c}}_{mn}$ 
is defined in \eqref{eta}. 

We have constructed the ${\cal N}=4$ instanton effective action 
in the $\Omega$-background from the ADHM method. 
In the next section, we will study the relation between the 
deformed instanton effective action and 
the D($-1$)-brane effective action for the 
D3/D($-1$)-brane system deformed in the R-R 3-form backgrounds. 

\section{D($-1$)-brane effective action in R-R backgrounds}
\subsection{Disk amplitudes in closed string backgrounds}
The
$\mathcal{N} = 4$ super Yang-Mills theory with gauge group $U(N)$ is 
realized as the low-energy effective theory of $N$ coincident D3-branes. 
The zero modes of open strings with both the end points 
on the D3-branes correspond to the $\mathcal{N} = 4$ vector 
multiplet. One can introduce instantons 
with the topological number $k$ as 
$k$ D$(-1)$-branes embedded into the D3-brane 
world-volume~\cite{Douglas:1995bn}. 
The zero modes of open strings 
at least one of whose end points is on the 
D$(-1)$-branes correspond to 
the ADHM moduli $a'_m, w_{\dot{\alpha}}, \bar{w}^{\dot{\alpha}},
\mathcal{M}^{\prime A}_{\alpha}, \mu^A, \bar{\mu}^A$ 
and the auxiliary variables $\chi_a, D^{\tilde{c}}, \bar{\psi}^A_{\dot{\alpha}}$~\cite{Witten:1994tz}.

The instanton effective action of the $\mathcal{N} = 4$ super Yang-Mills 
theory is also obtained as the low-energy effective action 
of the D$(-1)$-branes for the D3/D$(-1)$-brane system~\cite{Billo:2006jm}. 
The action is derived from the zero-slope limit 
$\alpha'\to 0$ of the open string disk amplitudes with the
boundary lying on the D$(-1)$-branes, where 
$\alpha'$ is the Regge slope parameter. 
The amplitudes include the vertex operators associated with the zero
modes. 
We summarize our conventions and notations for the calculations of 
the amplitudes in Appendix B. 
The vertex operators for the open string zero modes are found in
Table \ref{N4ADHM} in Appendix B.
We keep the D3-brane (Yang-Mills) coupling constant 
$g = (2\pi)^{\frac{1}{2}} g^{\frac{1}{2}}_s$
finite~\cite{Polchinski:1995mt}.
Then the zero-slope limit corresponds to the limit $g_0 \to \infty$, 
where $g_0 = (2\pi)^{-\frac{3}{2}}
\alpha^{\prime -1} g_s^{\frac{1}{2}}$ is the gauge coupling constant for 
the D$(-1)$-branes and $g_s$ is the string coupling constant. 
Some of the ADHM moduli in the vertex operators must be rescaled by $g_0$ 
to reproduce the field theory calculations \cite{Billo:2002hm}.

The D$(-1)$-brane action $S^{0}_{\mathrm{D}(-1)}$ 
which reproduces the amplitudes in the zero-slope limit
\cite{Billo:2002hm} is given by
\begin{eqnarray}
S^0_{\mathrm{D}(-1)}  
&=& \frac{2\pi^2}{\kappa} 
\mathrm{tr}_k 
\Big[
Y_{ma} Y_{ma} - X_{\dot{\alpha} a} \bar{X}^{\dot{\alpha}} {}_a
+ 2 Y_{ma} [\chi_a, a'_m] 
+ i D^{\tilde{c}} (\tau^{\tilde{c}})^{\dot{\alpha}} {}_{\dot{\beta}}
\left(
\bar{w}^{\dot{\beta}} w_{\dot{\alpha}} 
+ \bar{a}^{\prime \dot{\beta} \alpha} a'_{\alpha \dot{\alpha}}
\right)
\nonumber \\
& & \qquad \qquad 
- X_{\dot{\alpha} a} (\chi_a \bar{w}^{\dot{\alpha}} - 
\bar{w}^{\dot{\alpha}} \phi^0_a) - 
(w_{\dot{\alpha}} \chi_a - \phi^0_a w_{\dot{\alpha}}) 
\bar{X}^{\dot{\alpha}} {}_a
\nonumber \\
& & \qquad \qquad 
+ \frac{1}{2} (\bar{\Sigma}^a)_{AB} \bar{\mu}^A 
(- \chi_a \mu^B + \phi^0_a \mu^B)
- \frac{1}{2} (\bar{\Sigma}^a)_{AB} \mathcal{M}^{\prime \alpha A} 
\mathcal{M}'_{\alpha} {}^B \chi_a 
\nonumber \\
& & \qquad \qquad 
- i \bar{\psi}^{\dot{\alpha}} {}_A 
\left(
\bar{\mu}^A w_{\dot{\alpha}} 
+ \bar{w}_{\dot{\alpha}} \mu^A 
+ [\mathcal{M}^{\prime \alpha A}, a'_{\alpha \dot{\alpha}}]
\right)
\Big].
\end{eqnarray}
Here we have introduced the auxiliary fields 
$Y_{ma}, X_{\dot{\alpha} a}, \bar{X}_{\dot{\alpha}a }$ 
to disentangle the higher point interactions in the low-energy 
effective action \cite{Billo:2002hm}. 
After eliminating these auxiliary fields 
by using their equations of motion, the 
action $S^0_{\mathrm{D}(-1)}$ reduces to 
the one \eqref{eq:inst.eff} where all the deformation parameters are set 
to be zero.

We now study the deformation of the D$(-1)$-brane action in the R-R backgrounds. 
The R-R field strengths $\mathcal{F}_{M}, \mathcal{F}_{MNP},
\mathcal{F}_{MNPQR}$ can be combined into the bi-spinor form 
$\mathcal{F}^{\hat{\mathcal{A}} \hat{\mathcal{B}}}$, where $\hat{\mathcal{A}}, 
\hat{\mathcal{B}}$ are 16 component spinor indices 
in ten dimensions. 
Since the D3-branes break 
the $SO(10)$ Lorentz symmetry down to $SO(4) \times SO(6)$, 
the R-R backgrounds are decomposed into 
\begin{eqnarray}
\mathcal{F}^{\hat{\mathcal{A}} \hat{\mathcal{B}}} = 
(\mathcal{F}^{\alpha \beta AB}, \mathcal{F}^{\alpha} {}_{\dot{\beta}} 
{}^A {}_B, \mathcal{F}_{\dot{\alpha}} {}^{\beta} {}_A {}^B, 
\mathcal{F}_{\dot{\alpha} \dot{\beta} AB} ).
\label{eq:R-R_decomposition}
\end{eqnarray}
We consider the deformations of the effective action 
by the constant R-R backgrounds with the (S,A)-type 
$\mathcal{F}^{(\alpha \beta)[AB]}, \ 
\mathcal{F}_{(\dot{\alpha} \dot{\beta})[AB]}$ 
and the (A,S)-type 
$ \mathcal{F}^{[\alpha \beta](AB)}, \ \mathcal{F}_{[\dot{\alpha}
 \dot{\beta}](AB)} $ \cite{Ito:2007hy}. 
Here the round parentheses $(\cdot \cdot)$ denote symmetrization of
the indices and the square bracket $[\cdot \cdot]$ 
stands for anti-symmetrization. 
The (S,A)-type backgrounds 
%correspond to the R-R 3-form with 
have 
the index structure $\mathcal{F}_{mna}$ 
while the (A,S)-type backgrounds 
%correspond to 
have the index structure 
$\mathcal{F}_{abc}$.
In \cite{Ito:2009ac}, we have calculated the 
D$(-1)$-brane effective action 
in the presence of the constant self-dual R-R 
background of the (S,A)-type 
and have found that it agrees with the $\mathcal{N} = 4$ instanton 
effective action 
in the self-dual $\Omega$-background without the R-symmetry Wilson line 
by the identifications of $\Omega^{+}_{mna}$ with
$\mathcal{F}^{(\alpha\beta)[AB]}$.
%We have also found that the anti-self-dual (A,S)-type background
%$\mathcal{F}_{[\dot{\alpha}\dot{\beta}](AB)}$ 
%gives the deformation of the D$(-1)$-brane effective action 
%inducing the mass terms of the fermionic zero modes 
%$( \mathcal{M}'{}_{\alpha}^A , \mu^A , \bar{\mu}^A)$.
We have also observed that the (A,S)-type background
$\mathcal{F}_{[\dot{\alpha}\dot{\beta}](AB)}$ gives the holomorphic mass
terms for the fermionic ADHM moduli in the corresponding effective action.
In order to generalize this result to the non-(anti-)self-dual
$\Omega$-background with the Wilson line, 
we will consider both of the self-dual and the 
anti-self-dual parts of the (S,A)- 
and the (A,S)-type backgrounds. 
%{\bf since 
These backgrounds have the same tensor structures as those of the 
deformation parameters and give rise to the 
bilinear couplings with the massless fields in the D3/D$(-1)$ brane system. 
%}
We also introduce the constant NS-NS B-field background.
The low-energy effective theories of D-branes in the presence of the
constant NS-NS B-field can be described by gauge theories in the
noncommutative spacetime \cite{Chu:1998qz, Seiberg:1999vs}.
When one considers instantons in this background, 
the NS-NS B-field corresponds to the FI parameters 
in \eqref{eq:inst.eff}.

We study the effects of the closed string backgrounds in the 
D$(-1)$-brane effective theory by calculating the string disk amplitudes that 
contain the open and the closed string vertex operators.
The vertex operators for the closed string backgrounds are summarized in Table 
\ref{vertex_closed} in Appendix B.
As we discussed in \cite{Ito:2009ac}, 
we consider the zero-slope limit with finite 
$(2\pi \alpha')^{\frac{1}{2}} \mathcal{F}$. 
Here $\mathcal{F}$ represents the component of 
the (S,A)- and the (A,S)-type backgrounds. 

In order to cancel the overall factor $1/g_0^2$ 
of the disk amplitudes \eqref{disk_gen} in the zero-slope limit, 
the vertex operators for 
the following combinations of the fields need to be inserted 
in the disk amplitudes together with that for $\mathcal{F}$:
\begin{eqnarray}
\mu \bar{\mu}, \ Y a', \ 
\mathcal{M}' \mathcal{M}', \ w \bar{X}, \ \bar{w} X, \ 
X \bar{X}, \ w \bar{w}, \ Y Y, \ a' a'.
\label{comb}
\end{eqnarray}
By dimensional analysis and the conservation law of the charges 
associated with the spin operators and the twist fields, 
we find that the irreducible amplitudes that contain 
one vertex operator for the (S,A)- or the (A,S)-type backgrounds 
include only the 
vertex operators associated with the first five
combinations in \eqref{comb}.

In the following, we calculate 
the non-zero amplitudes that contain the closed string backgrounds.

\paragraph{Amplitudes with (S,A)-type backgrounds}
We first consider the amplitudes that contain the self-dual part of 
the (S,A)-type backgrounds~\cite{Ito:2009ac}. These have been evaluated as 
\begin{eqnarray}
\begin{aligned}
\langle \! \langle V^{(0)}_Y V^{(-1)}_{a'} V^{(-\frac{1}{2},-\frac{1}{2})}_{\mathcal{F}(+)}
\rangle \! \rangle &= 
\frac{2\pi^2}{\kappa} 
\mathrm{tr}_k 
\left[
- 4 \pi i 
\epsilon_{\beta \gamma} (\sigma^{mn})_{\alpha} {}^{\gamma} (\bar{\Sigma}^a)_{AB} 
Y_{ma} a'_n (2 \pi \alpha')^{\frac{1}{2}} \mathcal{F}^{(\alpha \beta) [AB]}
\right], \\ 
\langle \! \langle V^{(-\frac{1}{2})}_{\mathcal{M}'} V^{(-\frac{1}{2})}_{\mathcal{M}'} 
V^{(-\frac{1}{2},-\frac{1}{2})}_{\mathcal{F}(+)} \rangle \! \rangle &= 
\frac{2\pi^2}{\kappa} \mathrm{tr}_k 
\left[
- \pi (\bar{\Sigma}^a)_{AB} \mathcal{M}^{\prime A}_{\alpha} 
\mathcal{M}^{\prime B}_{\beta} (2 \pi \alpha')^{\frac{1}{2}} 
(\bar{\Sigma}^a)_{CD} \mathcal{F}^{(\alpha \beta) [CD]} 
\right].
\end{aligned}
\label{amp1}
 \end{eqnarray}
Next we consider the anti-self-dual part of the
(S,A)-type backgrounds, which are the new contributions. 
We find that the non-zero amplitudes are given by 
\begin{align}
\langle \! \langle 
V^{(0)}_{Y} V^{(-1)}_{a'} V^{(-\frac{1}{2},-\frac{1}{2})}_{\mathcal{F}(-)}
\rangle \! \rangle =& \frac{2\pi^2}{\kappa} 
\mathrm{tr}_k 
\left[
 4 \pi i \epsilon^{\dot{\beta} \dot{\gamma}}
(\bar{\sigma}^{mn})^{\dot{\alpha}} {}_{\dot{\gamma}} (\Sigma^a)^{AB}
Y_{ma} a'_n 
(2\pi \alpha')^{\frac{1}{2}} \mathcal{F}_{(\dot{\alpha} \dot{\beta})[AB]} 
\right], 
\label{SAamp1}
\\
\langle \! \langle V_{\bar{X}}^{(0)} V^{(-1)}_{w} 
V^{(-\frac{1}{2},-\frac{1}{2})}_{\mathcal{F}(-)} \rangle \! \rangle =& 
 \frac{2\pi^2}{\kappa} 
\mathrm{tr}_k 
\left[
- \pi X_{\dot{\alpha} a} 
(\bar{\sigma}_{mn})^{\dot{\alpha}} {}_{\dot{\beta}} 
\bar{w}^{\dot{\beta}} 
(2\pi 
\alpha')^{\frac{1}{2}} \mathcal{F}_{(\dot{\gamma} \dot{\delta})[AB]} 
\epsilon^{\dot{\delta} \dot{\kappa}}
(\bar{\sigma}^{mn})^{\dot{\gamma}} {}_{\dot{\kappa}} (\Sigma^a)^{AB}
\right], 
\label{SAamp2}
\nonumber \\
\\
\langle \! \langle V_{X}^{(0)} V^{(-1)}_{\bar{w}} 
V^{(-\frac{1}{2},-\frac{1}{2})}_{\mathcal{F}(-)} \rangle \! \rangle =& 
\frac{2\pi^2}{\kappa} \mathrm{tr}_k 
\left[
- \pi w_{\dot{\alpha}} 
(\bar{\sigma}_{mn})^{\dot{\alpha}} {}_{\dot{\beta}} 
\bar{X}^{\dot{\beta}}_a
(2\pi 
\alpha')^{\frac{1}{2}} \mathcal{F}_{(\dot{\gamma} \dot{\delta})[AB]} 
\epsilon^{\dot{\delta} \dot{\kappa}}
(\bar{\sigma}^{mn})^{\dot{\gamma}} {}_{\dot{\kappa}} (\Sigma^a)^{AB}
\right].
\nonumber \\
\label{SAamp3}
\end{align}
We leave the detailed calculations of these amplitudes to Appendix B.

\paragraph{Amplitudes with (A,S)-type backgrounds}
We next study the amplitudes that contain the (A,S)-type 
backgrounds. 
We find that all the amplitudes that involve the background 
$\mathcal{F}^{[\alpha \beta](AB)}$ vanish in the zero-slope limit.
On the other hand, some of the amplitudes 
that contain 
$\mathcal{F}_{[\dot{\alpha} \dot{\beta}](AB)}$ are non-zero. 
These amplitudes have been also calculated in~\cite{Ito:2009ac}, 
which are given by 
\begin{eqnarray}
\begin{aligned}
 & \langle \! \langle 
V^{(-\frac{1}{2})}_{\mathcal{M}'}
V^{(-\frac{1}{2})}_{\mathcal{M}'}
V^{(-\frac{1}{2},-\frac{1}{2})}_{\bar{\mathcal{F}}}
\rangle \! \rangle 
= \frac{2\pi^2}{\kappa} \mathrm{tr}_k 
\left[
2 \pi i \mathcal{M}^{\prime \alpha A} \mathcal{M}'_{\alpha} {}^B (2\pi 
\alpha')^{\frac{1}{2}} \mathcal{F}^{[\dot{\alpha} \dot{\beta}]} {}_{(AB)} 
\epsilon_{\dot{\alpha} \dot{\beta}}
\right], 
\\
 & 
\langle \! \langle 
V^{(-\frac{1}{2})}_{\bar{\mu}}
V^{(-\frac{1}{2})}_{\mu}
V^{(-\frac{1}{2},-\frac{1}{2})}_{\bar{\mathcal{F}}}
\rangle \! \rangle 
= \frac{2\pi^2}{\kappa} \mathrm{tr}_k 
\left[
2 \pi i \bar{\mu}^A \mu^B (2\pi 
\alpha')^{\frac{1}{2}} \mathcal{F}^{[\dot{\alpha} \dot{\beta}]} {}_{(AB)} 
\epsilon_{\dot{\alpha} \dot{\beta}}
\right].
\label{amp3}
\end{aligned}
\end{eqnarray}

\paragraph{Amplitudes with NS-NS B-field background}
We now calculate the disk amplitudes in the constant NS-NS B-field background. 
We first consider the amplitudes which do not contain any vertex operators 
for the R-R backgrounds but contain one vertex operator for 
the NS-NS B-field background and that for the open string zero modes. 
By dimensional analysis, 
the following amplitudes should be considered: 
\begin{equation}
\langle \! \langle 
V_D^{(0)} V_B^{(-1,-1)}
\rangle \! \rangle,
\quad 
\langle \! \langle 
V_{\chi}^{(0)} V_{\chi}^{(0)} V_B^{(-1,-1)}
\rangle \! \rangle, 
\quad 
\langle \! \langle 
V_{\phi}^{(0)} V_{\phi}^{(0)} V_B^{(-1,-1)}
\rangle \! \rangle, 
\quad 
\langle \! \langle 
V_{\chi}^{(0)} V_{\phi}^{(0)} V_B^{(-1,-1)}
\rangle \! \rangle.
\nonumber 
\end{equation}
The second, the third and the fourth ones vanish 
because the amplitudes are proportional to the 
factor $B_{mn} \delta^{mn}$. 
The first amplitude was evaluated in~\cite{Billo:2005fg}. 
The result is 
\begin{equation}
\langle \! \langle V_D^{(0)} V_B^{(-1,-1)} \rangle \! \rangle 
= \frac{2\pi^2}{\kappa} \mathrm{tr}_k [ i D_{\tilde{c}} \zeta^{\tilde{c}}],
\label{amp4}
\end{equation}
where we have defined $\zeta^{\tilde{c}} \equiv \bar{\eta}^{\tilde{c}}_{mn}
B^{mn}$. 

Next we consider the case where both the NS-NS B-field and the R-R 3-form 
backgrounds are turned on. 
Once these backgrounds are introduced simultaneously, 
there is a possibility of non-zero amplitudes containing both $B$ 
and $\mathcal{F}$.
By dimensional analysis, 
we find that the vertex operators for the open string 
zero modes in the amplitudes must not have $g_0$ dependence and 
the sum of their powers of 
$\alpha'$ must be 1/2. 
The only possible candidates for such vertex operators are
$V^{(-1)}_{\chi}$ and $V^{(-1)}_{\phi}$. 
Therefore we examine the following amplitudes:
\begin{eqnarray}
\langle \! \langle V^{(-1)}_{\chi} V^{(0,0)}_{B} V^{(-\frac{1}{2}, -\frac{1}{2})}_{\mathcal{F}} 
\rangle \! \rangle, 
\quad 
\langle \! \langle V^{(-1)}_{\phi} V^{(0,0)}_{B} V^{(-\frac{1}{2}, -\frac{1}{2})}_{\mathcal{F}} 
\rangle \! \rangle,
\label{amp_NSNS}
\end{eqnarray}
where $V_{\mathcal{F}}^{(-\frac{1}{2}, - \frac{1}{2})}$ is a vertex
operator for the (S,A)- or the (A,S)-type backgrounds.
Calculating these amplitudes are cumbersome since they contain 
the five-point world-sheet correlators. 
Instead 
we evaluate the amplitudes from the Wess-Zumino term of the
D$(-1)$-brane action in the 
NS-NS B-field and the R-R backgrounds \cite{Myers:1999ps}.
We find that the corresponding interaction term vanishes for the constant NS-NS
B-field and the (S,A)- or the (A,S)-type backgrounds. Then the amplitudes
\eqref{amp_NSNS} are zero\footnote{
The terms that contain the VEVs $\phi^0_a$ can not be derived from the
Wess-Zumino term in the effective action in \cite{Myers:1999ps}. 
However, the second amplitude in \eqref{amp_NSNS} vanishes
when the first amplitude is zero. 
This is because the structure of the vertex operator for $\chi_a$ and
$\phi^0_a$ is the same. 
}. 
The amplitudes that contain more than one R-R vertex operator are 
reducible or of higher order in $\alpha'$ and vanish in the zero-slope limit. 

\subsection{Deformed D$(-1)$-brane effective action}
We now consider the deformed D$(-1)$-brane effective action for the
D3/D$(-1)$-brane system. 
The amplitudes \eqref{amp1}--\eqref{amp4} 
are reproduced by the following effective action:
\begin{eqnarray}
S'_{\mathrm{D}(-1)} = S^0_{\mathrm{D}(-1)} + S_{\mathrm{closed}},
\end{eqnarray}
where 
\begin{eqnarray}
S_{\mathrm{closed}} &=&
\frac{2\pi^2}{\kappa}\,
\mathrm{tr}_{k}
\Big[
 2 Y_{ma} a'_n C^{mna} 
- \frac{1}{8} (\bar{\Sigma}^a)_{AB} 
\mathcal{M}^{\prime A}_{\alpha} \mathcal{M}_{\beta}^{\prime B} 
C^{mna} \epsilon^{\alpha \gamma} (\sigma_{mn})_{\gamma} {}^{\beta}
\nonumber \\
& & \qquad \qquad 
- \frac{1}{2} X_{\dot{\alpha} a} 
(\bar{\sigma}_{mn})^{\dot{\alpha}} {}_{\dot{\beta}}
\bar{w}^{\dot{\beta}} 
 C^{mna} 
- \frac{1}{2} w_{\dot{\alpha}} 
(\bar{\sigma}_{mn})^{\dot{\alpha} } {}_{\dot{\beta}} 
\bar{X}^{\dot{\beta}}_a C^{mna}
\nonumber \\
& & \qquad \qquad  
+ \left(
- 2 \bar{\mu}^A \mu^B - \mathcal{M}^{\prime \alpha A} \mathcal{M}^{\prime} 
{}_{\alpha} {}^B 
\right) \widetilde{m}_{AB}
- i D_{\tilde{c}} \zeta^{\tilde{c}}
\Big].
\end{eqnarray}
Here we have defined the deformation parameters $C^{mna}$ from 
the backgrounds $\mathcal{F}^{(\alpha \beta)[AB]}$, 
$\mathcal{F}_{(\dot{\alpha}  \dot{\beta})[AB]}$ 
and $\tilde{m}_{AB}$ from 
$\mathcal{F}_{[\dot{\alpha}\dot{\beta}](AB)}$
as 
\begin{align}
C^{mna} &= - 2 \pi (2\pi \alpha')^{\frac{1}{2}} 
\left[
\mathcal{F}^{(\alpha \beta)[AB]} 
\epsilon_{\beta \gamma} (\sigma^{mn})_{\alpha} {}^{\gamma}(\bar{\Sigma}^a)_{AB}
+
\mathcal{F}_{(\dot{\alpha}  \dot{\beta})[AB]} 
\epsilon^{\dot{\beta} \dot{\gamma}} (\bar{\sigma}^{mn})^{\dot{\alpha} }
{}_{\dot{\gamma}} (\Sigma^a)^{AB} 
\right], 
\label{def_parameters}
\\
\widetilde{m}_{AB} &= \pi i (2\pi \alpha')^{\frac{1}{2}} \mathcal{F}^{[\dot{\alpha} 
 \dot{\beta}]} {}_{(AB)} \epsilon_{\dot{\alpha} \dot{\beta}}.
\end{align}
After integrating out the auxiliary fields $Y_{ma}, X_{\dot{\alpha}a}, \bar{X}_{\dot{\alpha}a}$, 
we finally obtain the following 
effective action 
\begin{eqnarray}
S_{\mathrm{D}(-1)} \!\!\! &=& \!\!\! 
\frac{2\pi^2}{\kappa} \!
\mathrm{tr}_k 
\Big[
-\left(
[\chi_a, a'_m] + C_{mna} a^{\prime n}
\right)^2 
\nonumber \\
& & \qquad %\quad 
+ 
\left(
w_{\dot{\alpha}} \chi_a - \phi^0_a w_{\dot{\alpha}}
+ \frac{1}{2} C^{mna} (\bar{\sigma}_{mn})^{\dot{\gamma}} 
{}_{\dot{\alpha}} w_{\dot{\gamma}}
\right) 
\!\!
 \left(
\chi_a \bar{w}^{\dot{\alpha}} - \bar{w}^{\dot{\alpha}} \phi^0_a
+ \frac{1}{2} C^{pqa} (\bar{\sigma}_{pq})^{\dot{\alpha}} 
{}_{\dot{\beta}} \bar{w}^{\dot{\beta}}
\right)
\nonumber \\
& & \qquad %\quad 
+ \frac{1}{2} (\bar{\Sigma}^a)_{AB} \bar{\mu}^A 
(- \chi_a \mu^B + \phi^0_a \mu^B)
- \frac{1}{8} (\bar{\Sigma}^a)_{AB} 
\epsilon^{\alpha \gamma} (\sigma_{mn})_{\gamma} {}^{\beta} C^{mna}
\mathcal{M}'_{\alpha} {}^A \mathcal{M}'_{\beta} {}^B
\nonumber \\
& & 
\qquad %\quad 
- \frac{1}{2} (\bar{\Sigma}^a)_{AB} \mathcal{M}^{\prime \alpha A} 
\mathcal{M}'_{\alpha} {}^B \chi_a 
- 
\left(
2 \bar{\mu}^A \mu^B + \mathcal{M}^{\prime \alpha A} \mathcal{M}^{\prime} 
{}_{\alpha} {}^B
\right) \widetilde{m}_{AB}
\nonumber \\
& & \qquad %\qquad 
- i \bar{\psi}^{\dot{\alpha}}_A 
(\bar{\mu}^A w_{\dot{\alpha}} + \bar{w}_{\dot{\alpha}} \mu^A + 
[\mathcal{M}^{\prime \alpha A}, a'_{\alpha \dot{\alpha}}]) 
\nonumber \\
& & \qquad %\qquad 
+ i D^{\tilde{c}} (\tau^{\tilde{c}})^{\dot{\alpha}} {}_{\dot{\beta}} 
\left(\bar{w}^{\dot{\beta}} 
w_{\dot{\alpha}} + \bar{a}^{\prime \dot{\beta} \alpha} a'_{\alpha \dot{\alpha}}
- \frac{1}{2} (\tau^{\tilde{b}})^{\dot{\beta}} {}_{\dot{\alpha}} \zeta^{\tilde{b}}
\right)
\Big].
\label{Dinst_eff_action}
\end{eqnarray}
In this effective action, the (S,A)-type R-R 3-form backgrounds $C^{mna}$ 
give the mass terms for the bosonic moduli $(a'_m,
w_{\dot{\alpha}},\bar{w}^{\dot{\alpha}})$ and 
for the fermionic moduli $\mathcal{M}'{}^{A}_{\alpha}$ while the 
(A,S)-type backgrounds $\tilde{m}_{AB}$ give the mass terms for
the fermionic moduli $(\mathcal{M}'{}^{A}_{\alpha},\mu^A,\bar{\mu}^{A})$.
In fact, if we identify the deformation parameters $C_{mna}$ and 
$\widetilde{m}_{AB}$ with the $\Omega$-background and the 
mass parameters by $C^{mna} = - i \Omega^{mna}$, 
$\widetilde{m}_{AB} = m_{AB}$,
the low-energy effective action of 
the D$(-1)$-branes
in the 
constant NS-NS B-field, the (S,A)- and the (A,S)-type backgrounds 
coincides with 
the instanton effective action $S^{(0)}_{\mathrm{eff}}$
(\ref{eq:inst.eff}) in the 
$\Omega$-background with the R-symmetry Wilson line\footnote{A similar 
bi-spinor coupling has been studied for the matrix model in the pp-wave 
background \cite{Bonelli:2002mb}.}.

\section{Conclusions and Discussion}
In this paper we have studied the ${\cal N}=4$ super Yang-Mills theory 
deformed in the ten-dimensional $\Omega$-background with 
the R-symmetry Wilson line gauge field. 
We have obtained the deformed spacetime action in the 
general $\Omega$-background.
For the self-dual $\Omega$-background without the Wilson line we have 
shown that the spacetime action is invariant under the anti-chiral 
supersymmetry. 
For the general $\Omega$-background one expects that a part of the 
${\cal N}=4$ supersymmetry is preserved by choosing 
the Wilson line gauge field appropriately. 
In particular, we have shown that the action becomes the 
${\cal N}=2^*$ deformed one for the specific background, where the
hypermultiplet mass is introduced by the Wilson line.

We have constructed the supersymmetry 
transformations for the $\Omega$-deformed ${\cal N}=2^*$ theory 
explicitly. 
In the undeformed case, these transformations 
lead to the nilpotent fermionic charge where the
action is written as the exact form with respect to
the charge~\cite{Labastida:1997xk}.
In the deformed case, we can  show that 
the deformed ${\cal N}=2^*$ action is written in the exact form with respect
to the charge
defined by \eqref{eq:def2*.SUSY}.
The fermionic
charge is obtained by the topological twist of ${\cal N}=4$
supersymmetry, where one identifies the $SU(2)_R$ Lorentz group with 
the $SU(2)$ subgroup of the $SU(4)_I$ R-symmetry group.
One can consider three types of 
the different twists~\cite{Yamron:1988qc,Marcus:1995mq}. 
Among them, the half twist and the Vafa-Witten twist \cite{Vafa:1994tf} are
particularly interesting.
Since the ten-dimensional $\Omega$-background contains many deformation 
parameters,
it is interesting to explore the deformed supersymmetry of the
theory and  to construct the $\Omega$-deformed 
topologically twisted theories.

We have also studied the ADHM construction of instantons in the
$\Omega$-background.
Using the solutions to the equations of motion for the fields
at the leading order in the coupling constant, which are the same as the
ones in the self-dual $\Omega$-background, we got 
the deformed instanton effective action in the $\Omega$-background.
We have 
calculated the low-energy effective action of the D$(-1)$-branes for 
the D3/D$(-1)$ system in the R-R 3-form field strength backgrounds 
of the (S,A)- and the (A,S)-types and found that it agrees with the instanton 
effective action in the $\Omega$-background.
As in the case of the spacetime action, 
it is interesting to examine the deformed supersymmetry when the theory has 
nilpotent fermionic charges.

%{\bf 
The string theory calculation of the instanton effective action is 
rather straightforward compared to the ADHM method, where we need to 
solve the deformed equations of motion for 
the fields in the instanton background. 
We can apply the R-R 3-form field strength deformations to various 
D-brane systems. 
However the existence of the deformed supersymmetry 
transformations is not obvious in this approach. 
There is also a problem of the back reaction terms in the deformed 
D3-brane action, 
which is necessary to reproduce the 
deformed instanton effective action 
\cite{Ito:2009ac}.
%}

We note that the effective action \eqref{Dinst_eff_action} is different from 
the one discussed in \cite{Bruzzo:2002xf}
where the deformed action depends only on the 
$\epsilon$-parameters holomorphically. 
They also differ even in the case 
where all the deformation parameters vanish. 
The action in \cite{Bruzzo:2002xf} contains the quartic terms of 
$a'_m, \chi_a$ and the quadratic term of $D^{\tilde{c}}$ which are absent in
our action. This difference comes from the fact that 
when we consider the zero-slope limit of the amplitudes, 
we have rescaled some of the ADHM moduli by $g_0$ in the 
vertex operators as found in Table \ref{N4ADHM}. 
On the other hand, in \cite{Bruzzo:2002xf}, 
the authors studied 
the D$(-1)$-brane effective action without use of this rescaling. 
However both the effective actions 
provide 
the same instanton partition function \cite{Nekrasov:2003rj} 
(see \cite{Huang:2011qx} 
for the $\Omega$-deformed topological string amplitudes 
related to the ${\cal N}=2^*$ partition function). 
These problems will be discussed elsewhere.

\subsection*{Acknowledgments}
The work of K.~I. is supported in part by Grant-in-Aid
for Scientific Research from the Japan Ministry of Education, Culture,
Sports, Science and Technology.
The work of H.~N. is supported by Mid-career Researcher Program 
through the National Research Foundation of Korea(NRF) grant 
funded by the Korea government(MEST)(No. 2009-0084601). 
The work of T.~S. is supported by the Global
Center of Excellence Program by MEXT, Japan through the
``Nanoscience and Quantum Physics'' Project of the Tokyo
Institute of Technology, and by Iwanami Fujukai Foundation.
The work of S.~S. is supported by the Japan Society for the Promotion of Science (JSPS) Research Fellowship.

\begin{appendix}

\section{Sigma matrices in four and six dimensions} \label{sec:sigma}
In this appendix, 
 we present our conventions of the sigma matrices in four and six Euclidean dimensions.
The sigma matrices $\sigma^m_{\alpha\dot{\alpha}}$ and $\bar{\sigma}^{m \dot{\alpha}\alpha}$ 
 in four dimensions are defined 
 by 
\begin{align}
 \sigma^{m} &= \big( i \tau_1 , i \tau_2 , i \tau_3 , \mathbf{1}_2 \big) , & 
 \bar{\sigma}^m &= \bigl( -i \tau_1 , -i \tau_2 , -i \tau_3 , \mathbf{1}_2 \bigr) ,
\end{align}
 where 
 $\tau_{\tilde{c}} $ ($\tilde{c}=1,2,3$) 
 are the Pauli matrices
 and
 $\mathbf{1}_2$ denotes the $2 \times 2$ identity matrix.
We define the Lorentz generators $\sigma^{mn}$ and $\bar{\sigma}^{mn}$ by 
\begin{align}
 \sigma^{mn} &= \frac{1}{4} \big( \sigma^m \bar{\sigma}^n - \sigma^n \bar{\sigma}^m \big) , &
 \bar{\sigma}^{mn} &= \frac{1}{4} \big( \bar{\sigma}^m \sigma^n - \bar{\sigma}^n \sigma^m \big) .
\end{align}
They are related with the Pauli matrices by using the 't Hooft $\eta$-symbol:
\begin{align}
 \sigma_{mn} &= \frac{i}{2} \eta^{\tilde{c}}_{mn} \tau^{\tilde{c}} , &
 \bar{\sigma}_{mn} &= \frac{i}{2} \bar{\eta}^{\tilde{c}}_{mn} \tau^{\tilde{c}} .
 \label{eta}
\end{align}
The sigma matrices $(\Sigma_a )^{AB}$ and $(\bar{\Sigma}_a)_{AB}$ in six dimensions are defined 
 by 
\begin{align}
 \Sigma_a &= \big( \eta^3 , -i \bar{\eta}^3 , \eta^2 , -i \bar{\eta}^2 , \eta^1 , i \bar{\eta}^1 \big) , & 
 \bar{\Sigma}_a &= \bigl( - \eta^3 , -i \bar{\eta}^3 , - \eta^2 , -i \bar{\eta}^2 , - \eta^1 , i \bar{\eta}^1 \bigr) .
\end{align}
They are related by using the rank-4 $\epsilon$-symbol $\epsilon^{ABCD}$ with $\epsilon^{1234} =1$ as
\begin{align}
 (\Sigma_a )^{AB} = - \frac{1}{2} \epsilon^{ABCD} (\bar{\Sigma}_a)_{CD} .
 \label{eq:Sigma.relation}
\end{align}
The Lorentz generators $\Sigma_{ab}$ and $\bar{\Sigma}_{ab}$ in six dimensions are defined by 
\begin{align}
 \Sigma_{ab} &= \frac{1}{4} \big( \Sigma_a \bar{\Sigma}_b - \Sigma_b \bar{\Sigma}_a \big) , &
 \bar{\Sigma}_{ab} &= \frac{1}{4} \big( \bar{\Sigma}_a \Sigma_b - \bar{\Sigma}_b \Sigma_a \big) .
\end{align}

\section{Calculations of string disk amplitudes}
In this appendix, we calculate the string amplitudes 
\eqref{SAamp1}--\eqref{SAamp3} in Section 3.
We first summarize our notations and conventions of the 
world-sheet fields in type IIB superstring theory in 
ten-dimensional flat Euclidean spacetime. 
The world-sheet coordinates are denoted by $z, \bar{z}$. 
The fields $X^M (z, \bar{z})$, $\psi^M (z), \tilde{\psi}^M 
(\bar{z}), \ (M=1, \cdots, 10)$ are the bosonic and the fermionic string 
coordinates. The left moving fields %of the fields 
satisfy the free field OPEs which are given by 
\begin{eqnarray}
X^M (z) X^N (w) \sim \delta^{MN} \ln (z - w), \qquad 
\psi^M (z) \psi^N (w) \sim \delta^{MN} (z - w)^{-1}.
\end{eqnarray}
The right moving fields satisfy the similar OPEs. 
In the presence of parallel D3-branes, the $SO(10)$ Lorentz symmetry is 
broken down to $SO(4) \times SO(6)$ and the string coordinates are decomposed 
as $X^M = (X^m, X^{a+4}), \ \psi^M = (\psi^m, \psi^{a+4}) $, 
($m=1,\cdots,4$, $a=1,\cdots,6$), where 
$X^{m}$ spans the world-volume of the D3-branes and 
$X^{a+4}$ represents the transverse directions to the D3-branes.

The spin fields are defined by 
\begin{eqnarray}
\begin{aligned}
 & S^{\lambda} = e^{\lambda \phi \cdot e} c_{\lambda}, \quad 
 \phi \cdot e = \phi^i e_i \quad (i=1, \cdots, 5), \\
 & \lambda = \frac{1}{2} (\pm e_1 \pm 
e_2 \pm e_3 \pm e_4 \pm e_5) \equiv \lambda_i e_i,
\quad \lambda_i = \pm \frac{1}{2}, 
\end{aligned}
\label{spin_field}
\end{eqnarray}
where $e_i$ are the 
orthonormal basis in five dimensions, 
$c_{\lambda}$ is the cocycle factor and $\phi^i$ are the free bosons 
obtained by the bosonization of $\psi^M$ 
\cite{Kostelecky:1986xg}. 
The OPE of these free bosons is given by
\begin{eqnarray}
\phi^i (z) \phi^j (w) \sim \delta^{ij} \ln (z-w).
\end{eqnarray}
The weight vector $\lambda$ specifies the ten-dimensional 32 spinor components.
After the GSO projection, the spin fields corresponding to the weight vectors 
$\lambda$ that contain 
odd number of minus components survive. 
The ten-dimensional spin fields in the presence of the D3-branes 
are decomposed as 
\begin{eqnarray}
S^{\lambda} \to (S_{\alpha} S_A, S^{\dot{\alpha}} S^A),
\end{eqnarray}
where $\alpha, \dot{\alpha} = 1,2$ are the $SO(4)$ indices and 
$A=1,2,3,4$ are the $SO(6)$ spinor ($SU(4)$ (anti-)fundamental) indices. 
The explicit relation between the four-dimensional spinor indices and the spin 
states is found in~\cite{Ito:2010vx}. 

We also introduce the free boson field $\phi$ 
which is obtained by the bosonization of 
the superconformal ghost field and specifies the 
picture number of vertex operators 
~\cite{Friedan:1985ge}. The OPE is given by
\begin{equation}
\phi (z) \phi (w) \sim - \ln (z-w).
\end{equation}

We consider disk amplitudes containing closed and open
string vertex operators. A disk is realized as the upper half-plane 
parameterized by $z, \bar{z}$ and its boundary is the real axis
parametrized by $y$. 
We employ the doubling trick where the right moving fields of the closed
string vertex operators are
located on the lower-half plane and the left and 
the right moving fields are identified on the boundary.

The open string vertex operators corresponding to 
the ADHM moduli, the auxiliary variables, the auxiliary fields $Y_{ma}, X_{\dot{\alpha}a},
\bar{X}_{\dot{\alpha}a}$ and the VEVs of the scalar 
fields are summarized in Table \ref{N4ADHM}.
The zero modes of the D$(-1)$/D$(-1)$ open strings are in the adjoint
representation of $U(k)$ while 
those associated with the D3/D$(-1)$ strings are in the bi-fundamental
representations of $U(k)$ and $U(N)$. 
The $U(N)$ adjoint scalar VEVs are in the D3/D3 sector and are taken to be diagonal.
The powers of $\alpha'$ in the vertex operators are determined 
such that the zero modes have the canonical dimensions.
In order to reproduce the 
undeformed instanton effective action in the zero-slope limit, 
some of the zero modes should be rescaled by $g_0$~\cite{Billo:2002hm}. 
The fields $\Delta$ and $\bar{\Delta}$ are 
the twist fields which interchange the D3 and the D$(-1)$ 
boundaries~\cite{Dixon:1985jw,Hamidi:1986vh,Dixon:1986qv}. 
The twist fields appear as a pair of $\Delta$ and $\bar{\Delta}$ 
in the non-zero amplitudes.

\begin{table}[t]
\begin{center}
\begin{tabular}{|l||l|l|}
\hline
Brane sectors & Vertex operators & Zero modes \\
\hline \hline
D$(-1)$/D$(-1)$ &   $ \displaystyle V^{(-1)}_{a'} (y) = \pi (2 \pi 
 \alpha')^{\frac{1}{2}} g_0 a'_{m} \psi^{m}  e^{- \phi} 
 (y) / \sqrt{2}$ & ADHM moduli \\
            &   $ \displaystyle V_{\mathcal{M}}^{(- \frac{1}{2})} (y)= \pi (2 \pi \alpha')^{\frac{3}{4}} g_0
 \mathcal{M}^{\prime \alpha A}  S_{\alpha} S_A  e^{ - \frac{1}{2}  \phi} (y)$ &  \\
\cline{2-3}
            &   $ \displaystyle V_{\chi}^{(-1)} (y)= (2 \pi \alpha')^{\frac{1}{2}} 
 \chi_a \psi^{a+4}  e^{- \phi } (y) / \sqrt{2}$ &   Auxiliary variables\\
            &   $ \displaystyle V_{\bar{\psi}}^{(- \frac{1}{2})} (y)= 2 (2 \pi \alpha')^{\frac{3}{4}} 
 \bar{\psi}_{\dot{\alpha} A} S^{\dot{\alpha}}  S^A  e^{- \frac{1}{2} \phi}  (y) $  & \\
            &   $ \displaystyle V^{(0)}_D (y) = 
2 (2 \pi \alpha') D_c \bar{\eta}^c_{mn} \psi^{n} \psi^{m} (y) $ &   \\
\cline{2-3}
 & $ \displaystyle V^{(0)}_Y (y) = 
4 \pi (2 \pi \alpha') g_0 Y_{ma} \psi^m \psi^{a+4} (y)$ & Auxiliary fields \\
\hline
D3/D$(-1)$ &   $\displaystyle V_{w}^{(-1)} (y) = 
\pi (2 \pi \alpha')^{\frac{1}{2}} g_0 w_{\dot{\alpha}} 
\Delta S^{\dot{\alpha}} e^{- \phi } (y) / 2$ & ADHM moduli \\
         &   $\displaystyle V^{(-1)}_{\bar{w}} (y) = \pi (2 \pi 
 \alpha')^{\frac{1}{2}} g_0 \bar{w}_{\dot{\alpha}} \bar{\Delta} 
 S^{\dot{\alpha}} e^{- \phi} (y) / 2$ &  \\
        &   $\displaystyle V_{\mu}^{(-\frac{1}{2})} (y)= \pi (2 \pi \alpha')^{\frac{3}{4}} g_0 
\mu^A \Delta S_A e^{- \frac{1}{2} \phi } (y) $ &  \\
        &   $\displaystyle V_{\bar{\mu}}^{(-\frac{1}{2})} (y)= \pi (2 \pi \alpha')^{\frac{3}{4}} g_0 
\bar{\mu}^A \bar{\Delta} S_A e^{- \frac{1}{2} \phi } (y) $ &  \\
\cline{2-3}
 & $\displaystyle V_X^{(0)} = 2 \sqrt{2} \pi (2 \pi \alpha') g_0
	 X_{\dot{\alpha} a} \Delta S^{\dot{\alpha}} \psi^{a+4} (y)$ & Auxiliary
		 fields \\
 & $\displaystyle V_{\bar{X}}^{(0)} = 2 \sqrt{2} \pi (2 \pi \alpha') g_0
	 \bar{X}_{\dot{\alpha} a} \bar{\Delta} S^{\dot{\alpha}} \psi^{a+4} (y)$
	 & \\
\hline
D3/D3& $ \displaystyle \frac{}{} V^{(-1)}_{\phi} (y) = 
(2 \pi \alpha')^{\frac{1}{2}} \phi^0_a \psi^{a+4} e^{- 
\phi} (y) / \sqrt{2}$ & Scalar VEVs \\
\hline
\end{tabular}
\caption{Vertex operators for the open string zero modes
in the D-brane sectors. 
We denote the vertex operator for a field $\Psi$ in the $q$-picture
by $V_{\Psi}^{(q)}$. We omit the normal ordering symbol.
}
\label{N4ADHM}
\end{center}
\end{table}

The vertex operators for the NS-NS B-field, the (S,A)- and the (A,S)-type
backgrounds are given in Table \ref{vertex_closed}. As mentioned in
Section 3, we consider the scaling such that $(2\pi
\alpha')^{\frac{1}{2}} \mathcal{F}$ is finite in the zero-slope limit. 
Here $\mathcal{F}$ is the component of the (S,A)- and 
the (A,S)-type backgrounds. 
\begin{table}[t]
\begin{center}
\begin{tabular}{|l||l|}
\hline
Backgrounds & Vertex operators \\
\hline \hline
NS-NS B-field & $ \displaystyle \frac{}{} V^{(-1,-1)}_{B} (z,\bar{z})= 
- \tfrac{i}{4} \pi^2 (2 \pi \alpha') g_0^{2} B_{mn} \psi^m e^{- 
 \phi} (z) \psi^n e^{-  \phi} (\bar{z}) $ \\
\hline
(S,A)-type R-R 3-form & 
$ \displaystyle \frac{}{} V^{(-\frac{1}{2},-\frac{1}{2})}_{\mathcal{F}(+)} (z,\bar{z}) = 
 (2 \pi \alpha') \mathcal{F}^{(\alpha \beta) [AB]} S_{\alpha} S_A 
e^{-\frac{1}{2} \phi} (z) S_{\beta} S_B e^{- \frac{1}{2} \phi} (\bar{z}) $ \\
\cline{2-2}
           & 
$ \displaystyle \frac{}{} V^{(-\frac{1}{2},-\frac{1}{2})}_{\mathcal{F} (-)} 
(z, \bar{z}) = 
(2 \pi \alpha') \mathcal{F}_{(\dot{\alpha} \dot{\beta}) [AB]} S^{\dot{\alpha}} S^A 
e^{-\frac{1}{2} \phi} (z) S^{\dot{\beta}} S^B e^{- \frac{1}{2} \phi} (\bar{z}) $ \\
\hline
(A,S)-type R-R 3-form & 
$ \displaystyle \frac{}{} V^{(-\frac{1}{2},-\frac{1}{2})}_{\widetilde{\mathcal{F}}} (z,\bar{z}) = 
 (2 \pi \alpha') \mathcal{F}^{[\alpha \beta] (AB)} S_{\alpha} S_A 
e^{-\frac{1}{2} \phi} (z) S_{\beta} S_B e^{- \frac{1}{2} \phi} (\bar{z}) $ \\
\cline{2-2}
           & 
$ \displaystyle \frac{}{} V^{(-\frac{1}{2},-\frac{1}{2})}_{\bar{\mathcal{F}}} 
(z, \bar{z}) = 
(2 \pi \alpha') \mathcal{F}_{[\dot{\alpha} \dot{\beta}] (AB)} S^{\dot{\alpha}} S^A 
e^{-\frac{1}{2} \phi} (z) S^{\dot{\beta}} S^B e^{- \frac{1}{2} \phi} (\bar{z}) $ \\
\hline
\end{tabular}
\caption{Vertex operators for the closed string backgrounds. 
The factor $g_0^2$ in the B-field vertex operator must be
introduced to reproduce the FI terms in \eqref{amp4}.}
\label{vertex_closed}
\end{center}
\end{table}

The disk amplitude which contains $n_o$ open string vertex operators 
$V_{\Psi_i}^{(q_i)} (y_i)$ and $n_c$ closed string vertex operators 
$V_{C_i}^{(r_i, \tilde{r}_i)}$ is given by 
\begin{equation}
\langle \! \langle V^{(q_1)}_{\Psi_1}\cdots V^{(r_1,\tilde{r}_1)}_{C_1}
\cdots \rangle \! \rangle
= C_0 \int \frac{\prod_{i=1}^{n_o} dy_i \prod_{j=1}^{n_c}
d z_j d \bar{z}_j}{dV_{CKG}}
\langle V^{(q_1)}_{\Psi_1} (y_1) \cdots V^{(r_1,\tilde{r}_1)}_{C_1} 
(z_1,\bar{z_1}) \cdots \rangle.
\label{disk_gen}
\end{equation}
Here $C_0$ is the disk normalization factor which is given by 
\begin{equation}
 C_0 = \frac{1}{2\pi^2 \alpha^{\prime 2}} \frac{1}{\kappa g_0^2}.
\end{equation}
The factor $dV_{CKG}$ is the $SL(2,\mathbb{R})$-invariant volume element to 
fix three positions $x_1$, $x_2$ and $x_3$ among $y_i$, $z_j$ and 
$\bar{z}_j$'s. This is given by
\begin{equation}
 d V_{CKG}={d x_1 d x_2 d x_3\over
(x_1-x_2) (x_2-x_3) (x_3-x_1)}.
\end{equation}
We fix one position of an open string vertex operator 
to $y_1 \to \infty$ and the other positions of a closed string vertex operator 
to $z = i, \bar{z} = -i$.
Note that in the disk amplitude (\ref{disk_gen}) the sum of the 
picture numbers must be $-2$. 

In the following, we calculate the amplitudes 
\eqref{SAamp1}--\eqref{SAamp3}.
Using the formula \eqref{disk_gen}, the amplitude \eqref{SAamp1} is 
evaluated as 
\begin{eqnarray}
\langle \! \langle 
V^{(0)}_Y V^{(-1)}_{a'} V^{(- \frac{1}{2},-\frac{1}{2})}_{\mathcal{F} (-)}
\rangle \! \rangle
&=& 
\frac{1}{2\pi^2 \alpha^{\prime 2}} \frac{1}{\kappa g_0^2} 
(2\pi \alpha')^2 g_0^2 
\left(
\frac{4\pi^2}{\sqrt{2}}
\right)
\mathrm{tr}_k
\left[
Y_{ma} a'_n (2 \pi \alpha')^{\frac{1}{2}} \mathcal{F}_{(\dot{\alpha} \dot{\beta})[AB]}
\right]
\nonumber \\
& & \times \int^{y_1}_{- \infty} \! d y_2 \ (y_1 - z) (y_1 - \bar{z}) (z 
- \bar{z}) 
\nonumber \\
& & \times 
\langle e^{- \phi (y_2)} e^{- \frac{1}{2} \phi (z)} e^{- \frac{1}{2} 
\phi (\bar{z})} \rangle
\nonumber \\
& & \times 
\langle 
\psi^m \psi^{a+4} (y_1) 
\psi^n (y_2) S^{\dot{\alpha}} (z) S^A (z) S^{\dot{\beta}} (\bar{z}) S^B (\bar{z})
\rangle.
\end{eqnarray}
The correlators involving the world-sheet free fields are evaluated by
the formulas in~\cite{Kostelecky:1986xg}.
After fixing three positions of the vertices and performing the integration 
of $y_{2}$, 
%the position of the vertex operator, 
we obtain 
\begin{eqnarray}
\langle \! \langle 
V^{(0)}_Y V^{(-1)}_{a'} V^{(-\frac{1}{2},-\frac{1}{2})}_{\mathcal{F}(-)}
\rangle \! \rangle
= - \frac{2\pi^2}{\kappa} 
\mathrm{tr}_k 
\left[
2 Y_{ma} a'_n C^{(-)mna}
\right],
\end{eqnarray}
where the cocycle phase factor becomes $1$ and 
$C^{(-)mna}$ is the anti-self-dual part of the deformation
parameter \eqref{def_parameters}.

The amplitude \eqref{SAamp2} is given by 
\begin{eqnarray}
\langle \! \langle V^{(0)}_X V^{(-1)}_{\bar{w}} 
V^{(-\frac{1}{2},-\frac{1}{2})}_{\mathcal{F}(-)} \rangle \! \rangle 
&=& \frac{1}{2 \pi^2 \alpha^{\prime 2}} \frac{1}{\kappa g_0^2} 
(2 \sqrt{2} \pi) \frac{\pi}{2} \mathrm{tr}_k 
\left[
X_{\dot{\alpha} a} \bar{w}_{\dot{\beta}} (2 \pi \alpha')^{\frac{1}{2}} 
\mathcal{F}_{(\dot{\gamma} \dot{\delta}) [AB]} 
\right]
\nonumber \\
& & \times \int^{y_1}_{- \infty} \! dy_2 
\ (y_1 - z) (y_1 - \bar{z}) (z - \bar{z}) 
\times 
\langle e^{- \phi (y_2)} e^{- \frac{1}{2} \phi (z)} e^{- \frac{1}{2} 
\phi (\bar{z})} \rangle 
\nonumber \\
& & \times
\langle \Delta (y_1) \bar{\Delta} (y_2) \rangle 
\langle S^{\dot{\alpha}} (y_1) S^{\dot{\beta}} (y_2) S^{\dot{\gamma}} (z) 
S^{\dot{\delta}} (\bar{z}) \rangle 
\langle \psi^{a+4} (y_1) S^A (z) S^B (\bar{z}) \rangle. 
\nonumber \\
\end{eqnarray}
Again, the correlators which contain the ghost and the spin fields are evaluated using the
formulas in~\cite{Kostelecky:1986xg}. 
The twist field correlator is evaluated as 
~\cite{Dixon:1985jw,Hamidi:1986vh,Dixon:1986qv}
\begin{eqnarray}
\langle \Delta (y_1) \bar{\Delta} (y_2) \rangle = (y_1 - y_2)^{-\frac{1}{2}}.
\end{eqnarray}
Then the result is 
\begin{eqnarray}
\langle \! \langle V^{(0)}_X V^{(-1)}_{\bar{w}} 
V^{(-\frac{1}{2},-\frac{1}{2})}_{\mathcal{F} (-)} \rangle \! \rangle 
= \frac{2\pi^2}{\kappa} 
\mathrm{tr}_k 
\left[
\frac{1}{2} X_{\dot{\alpha} a} 
(\bar{\sigma}_{mn})^{\dot{\alpha} } {}_{\dot{\beta}} 
\bar{w}^{\dot{\beta}} 
C^{mna}
\right],
\end{eqnarray}
where the cocycle phase factor $-i$ has been included.

The amplitude \eqref{SAamp3} is evaluated in the same manner.
The result is 
\begin{eqnarray}
\langle \! \langle 
V^{(-1)}_w V^{(0)}_{\bar{X}} V^{(-\frac{1}{2},-\frac{1}{2})}_{\mathcal{F} (-)}
\rangle \! \rangle 
= \frac{2\pi^2}{\kappa} \mathrm{tr}_k 
\left[
\frac{1}{2} w_{\dot{\alpha}}
(\bar{\sigma}_{mn})^{\dot{\alpha} } {}_{\dot{\beta}}
\bar{X}^{\dot{\beta}a} 
C^{mna}
\right].
\end{eqnarray}

\end{appendix}


\begin{thebibliography}{99}

\bibitem{Moore:1997dj}
  G.~W.~Moore, N.~Nekrasov and S.~Shatashvili,
  Commun.\ Math.\ Phys.\  {\bf 209} (2000) 97
  [arXiv:hep-th/9712241].



\bibitem{Nekrasov:2002qd}
  N.~A.~Nekrasov,
  Adv.\ Theor.\ Math.\ Phys.\  {\bf 7} (2004) 831
  [arXiv:hep-th/0206161].


\bibitem{LoMaNe}
A.~S.~Losev, A.~Marshakov and N.~A.~Nekrasov,
arXiv:hep-th/0302191.

\bibitem{Nekrasov:2003rj}
  N.~Nekrasov and A.~Okounkov,
  arXiv:hep-th/0306238.


\bibitem{AnGaNaTa}
I.~Antoniadis, E.~Gava, K.~S.~Narain and T.~R.~Taylor,
Nucl.\ Phys.\ B {\bf 413} (1994) 162
[arXiv:hep-th/9307158].

\bibitem{BeCeOoVa}
M.~Bershadsky, S.~Cecotti, H.~Ooguri and C.~Vafa,
Commun.\ Math.\ Phys.\ {\bf 165} (1994) 311
[arXiv:hep-th/9309140].


\bibitem{Awata:2005fa}
  H.~Awata and H.~Kanno,
  JHEP {\bf 0505} (2005) 039
  [arXiv:hep-th/0502061].

\bibitem{Iqbal:2007ii}
  A.~Iqbal, C.~Kozcaz and C.~Vafa,
  JHEP {\bf 0910} (2009) 069
  [arXiv:hep-th/0701156].

\bibitem{Huang:2010kf}
  M.~x.~Huang and A.~Klemm,
  arXiv:1009.1126 [hep-th].

\bibitem{Antoniadis:2010iq}
  I.~Antoniadis, S.~Hohenegger, K.~S.~Narain, T.~R.~Taylor,
  Nucl.\ Phys.\  {\bf B838 } (2010)  253
  [arXiv:1003.2832 [hep-th]].

\bibitem{Nakayama:2011be}
  Y.~Nakayama and H.~Ooguri,
  arXiv:1106.5503 [hep-th].


\bibitem{Billo:2006jm}
  M.~Billo, M.~Frau, F.~Fucito and A.~Lerda,
  JHEP {\bf 0611} (2006) 012
  [arXiv:hep-th/0606013].

\bibitem{Ito:2010vx}
  K.~Ito, H.~Nakajima, T.~Saka and S.~Sasaki,
  JHEP {\bf 1011} (2010) 093
  [arXiv:1009.1212 [hep-th]].

\bibitem{Jafferis:2007sg}
  D.~L.~Jafferis,
  arXiv:0705.2250 [hep-th].

\bibitem{Awata:2009dd}
  H.~Awata and H.~Kanno,
  JHEP {\bf 0907} (2009) 076
  [arXiv:0905.0184 [hep-th]].

\bibitem{Ne-M}
  N.~Nekrasov,
  Japan. J. Math. {\bf 4} (2009) 63.

\bibitem{Billo:2009di}
  M.~Billo, L.~Ferro, M.~Frau, L.~Gallot, A.~Lerda and I.~Pesando,
  JHEP {\bf 0907} (2009) 092
  [arXiv:0905.4586 [hep-th]].

\bibitem{Fucito:2009rs}
  F.~Fucito, J.~F.~Morales and R.~Poghossian,
  JHEP {\bf 0910} (2009) 041
  [arXiv:0906.3802 [hep-th]].

\bibitem{Billo:2010mg}
  M.~Billo, L.~Gallot, A.~Lerda and I.~Pesando,
  JHEP {\bf 1011} (2010) 041
  [arXiv:1008.5240 [hep-th]].



\bibitem{Ito:2009ac}
  K.~Ito, H.~Nakajima, T.~Saka and S.~Sasaki,
  JHEP {\bf 0910} (2009) 028
  [arXiv:0908.4339 [hep-th]].

\bibitem{Hollowood:2002zv}
  T.~J.~Hollowood,
  Nucl.\ Phys.\  B {\bf 639} (2002) 66
  [arXiv:hep-th/0202197].


\bibitem{Bruzzo:2002xf}
  U.~Bruzzo, F.~Fucito, J.~F.~Morales and A.~Tanzini,
  JHEP {\bf 0305} (2003) 054
  [arXiv:hep-th/0211108].

\bibitem{Brink:1976bc}
  L.~Brink, J.~H.~Schwarz, J.~Scherk,
  Nucl.\ Phys.\  {\bf B121 } (1977)  77.

\bibitem{Witten:1988ze}
  E.~Witten,
  Commun.\ Math.\ Phys.\  {\bf 117} (1988) 353.




\bibitem{Affleck:1980mp}
  I.~Affleck,
  Nucl.\ Phys.\ {\bf B191} (1981) 429.

\bibitem{AtHiDrMa}
M.~F.~Atiyah, N.~J.~Hitchin, V.~G.~Drinfeld and Yu.~I.~Manin,
Phys.\ Lett.\ A {\bf 65} (1978) 185.

\bibitem{DoHoKhMa}
N.~Dorey, T.~J.~Hollowood, V.~V.~Khoze and M.~P.~Mattis,
Phys.\ Rept.\ {\bf 371} (2002) 231
[arXiv:hep-th/0206063].


\bibitem{NeSc}
N.~Nekrasov and A.~S.~Schwarz,
Commun.\ Math.\ Phys.\ {\bf 198} (1998) 689
[arXiv:hep-th/9802068].

\bibitem{Douglas:1995bn}
  M.~R.~Douglas,
  arXiv:hep-th/9512077.


\bibitem{Witten:1994tz}
  E.~Witten,
  J.\ Geom.\ Phys.\  {\bf 15} (1995) 251
  [arXiv:hep-th/9410052].

\bibitem{Polchinski:1995mt}
  J.~Polchinski,
  Phys.\ Rev.\ Lett.\  {\bf 75} (1995) 4724
  [arXiv:hep-th/9510017].


\bibitem{Billo:2002hm}
  M.~Billo, M.~Frau, I.~Pesando, F.~Fucito, A.~Lerda, A.~Liccardo,
  JHEP {\bf 0302} (2003) 045
  [arXiv:hep-th/0211250].

\bibitem{Ito:2007hy}
  K.~Ito, H.~Nakajima, S.~Sasaki,
  JHEP {\bf 0707} (2007) 068
  [arXiv:0705.3532 [hep-th]].

\bibitem{Chu:1998qz}
  C.~-S.~Chu, P.~-M.~Ho,
  Nucl.\ Phys.\  {\bf B550} (1999) 151
  [arXiv:hep-th/9812219].

\bibitem{Seiberg:1999vs}
  N.~Seiberg, E.~Witten,
  JHEP {\bf 9909} (1999) 032
  [arXiv:hep-th/9908142].

\bibitem{Billo:2005fg}
  M.~Billo, M.~Frau, S.~Sciuto, G.~Vallone, A.~Lerda,
  JHEP {\bf 0605} (2006) 069
  [arXiv:hep-th/0511036].

\bibitem{Myers:1999ps}
  R.~C.~Myers,
  JHEP {\bf 9912} (1999) 022
  [arXiv:hep-th/9910053].

\bibitem{Bonelli:2002mb}
  G.~Bonelli,
  JHEP {\bf 0208} (2002) 022
  [arXiv:hep-th/0205213].


\bibitem{Labastida:1997xk}
  J.~M.~F.~Labastida and C.~Lozano,
  Nucl.\ Phys.\  B {\bf 518} (1998) 37
  [arXiv:hep-th/9711132].

\bibitem{Yamron:1988qc}
  J.~P.~Yamron,
  Phys.\ Lett.\  {\bf B213} (1988) 325.

\bibitem{Marcus:1995mq}
  N.~Marcus,
  Nucl.\ Phys.\ {\bf B452} (1995) 331
  [arXiv:hep-th/9506002].

\bibitem{Vafa:1994tf}
   C.~Vafa and E.~Witten,
   Nucl.\ Phys.\ {\bf B431} (1994) 3
   [arXiv:hep-th/9408074].




\bibitem{Kostelecky:1986xg}
  V.~A.~Kostelecky, O.~Lechtenfeld, W.~Lerche, S.~Samuel, S.~Watamura,
  Nucl.\ Phys.\  {\bf B288} (1987) 173.

\bibitem{Friedan:1985ge}
  D.~Friedan, E.~J.~Martinec, S.~H.~Shenker,
  Nucl.\ Phys.\  {\bf B271} (1986) 93.

\bibitem{Dixon:1985jw}
  L.~J.~Dixon, J.~A.~Harvey, C.~Vafa, E.~Witten,
  Nucl.\ Phys.\  {\bf B261} (1985) 678.

\bibitem{Hamidi:1986vh}
  S.~Hamidi, C.~Vafa,
  Nucl.\ Phys.\  {\bf B279} (1987) 465.

\bibitem{Dixon:1986qv}
  L.~J.~Dixon, D.~Friedan, E.~J.~Martinec, S.~H.~Shenker,
  Nucl.\ Phys.\  {\bf B282} (1987) 13.


\bibitem{Huang:2011qx} 
  M.~-x.~Huang, A.~-K.~Kashani-Poor and A.~Klemm,
  arXiv:1109.5728 [hep-th].
\end{thebibliography}
\end{document}